\documentclass[11pt]{article}

\usepackage{epsfig}
\usepackage{amsmath}
\usepackage{amssymb}
\usepackage{xypic}

\newcommand{\mbf}[1]{{\boldsymbol {#1} }}

\renewcommand{\thefootnote}{\fnsymbol{footnote}}
\newcommand{\newsection}{\setcounter{equation}{0}\section}

\def\appendix#1{\addtocounter{section}{1}\setcounter{equation}{0}
\renewcommand{\thesection}{\Alph{section}}
\section*{Appendix \thesection\protect\indent \parbox[t]{11.715cm} {#1}}
\addcontentsline{toc}{section}{Appendix \thesection\ \ \ #1} }
\newcommand{\complex}{{\mathbb C}} 
\newcommand{\zed}{{\mathbb Z}} 
\newcommand{\nat}{{\mathbb N}} 
\newcommand{\real}{{\mathbb R}} 
\newcommand{\zeds}{{\mathbb Z}} 
\newcommand{\torus}{{\mathbb T}}
\newcommand{\ann}{{\mathbb A}}
\newcommand{\disk}{{\mathbb B}}
\newcommand{\elliptic}{{\mathbb E}}
\newcommand{\id}{{1\!\!1}} 
\newcommand{\sympl}{{\sf J}}
\newcommand{\homo}{{\sf M}}
\newcommand{\pull}{{\sf H}}
\newcommand{\sbullet}{\,{\scriptstyle\bullet}\,}

\newif\ifold             \oldtrue            

\def\nn{\nonumber}

\newcommand{\Tr}{\:{\rm Tr}\,}

\def\e{{\,\rm e}\,}
\newcommand{\ii}{\,{\rm i}\,}
\newcommand{\dd}{{\rm d}}

\def\benum{\begin{enumerate}}
\def\eenum{\end{enumerate}}
\def\be{\begin{equation}}
\def\ee{\end{equation}}
\def\bea{\begin{eqnarray}}
\def\eea{\end{eqnarray}}
\def\bd{\begin{displaymath}}
\def\ed{\end{displaymath}}

\newcommand{\beq}{\begin{equation}}
\newcommand{\eeq}{\end{equation}}

\newcommand{\z}{\zeta}

\newtheorem{Poincare's Normal Form Lemma.}{Poincare's Normal Form Lemma.}
\newtheorem{Riemann Hurwitz}{Theorem, The Riemann Hurwitz Formula}

\allowdisplaybreaks

\begin{document}
\begin{titlepage}
\begin{flushright}

\baselineskip=12pt

HWM--06--4\\
EMPG--06--01\\
hep--th/0601220\\
\hfill{ }\\
January 2006
\end{flushright}

\begin{center}

\baselineskip=24pt

\vspace{2cm}

{\Large\bf Two-Loop String Theory on Null Compactifications}

\baselineskip=14pt

\vspace{2cm}

{\bf Henry C.D. Cove} and {\bf Richard J. Szabo}
\\[4mm]
{\it Department of Mathematics}\\ and\\ {\it Maxwell Institute for
  Mathematical Sciences\\Heriot-Watt University\\ Colin Maclaurin
  Building, Riccarton, Edinburgh EH14 4AS, U.K.}
\\{\tt henry@ma.hw.ac.uk} , {\tt R.J.Szabo@ma.hw.ac.uk}
\\[50mm]

\end{center}

\begin{abstract}

We compute the two-loop contributions to the free energy in the null
compactification of perturbative string theory at finite
temperature. The cases of bosonic, Type~II and heterotic strings are
all treated. The calculation exploits an explicit reductive
parametrization of the moduli space of infinite-momentum frame string
worldsheets in terms of branched cover instantons. Various arithmetic
and physical properties of the instanton sums are
described. Applications to symmetric product orbifold conformal field
theories and to the matrix string theory conjecture are also briefly
discussed.

\end{abstract}

\end{titlepage}

\setcounter{page}{2}
\renewcommand{\thefootnote}{\arabic{footnote}}
\setcounter{footnote}{0}

\newsection{Introduction and Summary\label{Intro}}

Compactifications of string theory along a dimension which is
light-like, rather than space-like, are of interest for a variety of
reasons (See~\cite{Sem1} for reviews of some of the issues addressed
in the following). A light-like circle can be gotten from a space-like
one by an infinite Lorentz boost~\cite{Seiberg1}. By T-duality, for
any light-like compactification radius $R$ the energy spectrum of the
rest frame states coincides with that of the uncompactified string
theory. These compactifications thereby probe the T-dual string theory
in a regime wherein long fundamental string states wrap an almost
infinite compact direction. In discrete light-cone quantization (DLCQ)
on flat ten-dimensional spacetime, the momentum $p^+$ along the compact
null direction $x^-$ is quantized as
\beq
p^+=\frac NR
\label{p+quant}\eeq
with $N\in\nat$, while the light-cone energy $p^-$ is determined by
the mass-shell relation
\beq
p^-=\frac1{p^+}\,\bigl(L_0+\overline{L}{}_0\bigr)=
\frac1{p^+}\,H \ .
\label{massshell}\eeq
Analysis of the Hilbert space shows~\cite{DMVV1,Dijk1,FM1} that free
second-quantized Type~IIA superstring theory is naturally equivalent
to a free superconformal sigma-model on the symmetric product orbifold
\beq
{\rm Sym}^N\big(\real^8\big)=\real^{8N}/S_N \ ,
\label{symprodorb}\eeq
in that the corresponding conformal field theory vacuum amplitudes
coincide in the free string infrared limit $g_s\to0$.

This equivalence may be given a geometric interpretation by
introducing a finite temperature~\cite{GS1,GOPS1}. This is done by further
compactifying Euclidean time so that two target space directions are
compactified on a torus $\torus_\tau^2$ of a particular modulus
$\tau$. For the present discussion, the thermodynamic partition
function is simply regarded as a generating function for the energy
spectrum of free string theory and thermal instabilities such as the
gravitational Jeans instability or the stringy Hagedorn transition
will be ignored. On the Type~IIA side, the one-loop free energy is
given by a sum over unramified coverings of the torus $\torus^2_\tau$
of degree $N$~\cite{GS1,GOPS1}. On the superconformal field theory
side, the partition function on $\torus_\tau^2$ is given by a sum over
twisted sectors imposing $S_N$-twisted boundary conditions on the
string embedding fields. An extra summation over elements of $S_N$ is
required to define a projection onto the $S_N$-invariant subspace of
the Hilbert space, resulting in a sum over commuting pairs of permutations
assuring that the twists in time and space directions commute. The
twisted sectors have a natural interpretation in terms of ``long''
strings formed from ``short'' fundamental string bits. The partition
function from $N$ fundamental single strings are combined together to
give the partition function of one long string with a modified modular
parameter, i.e. the worldsheet of the long strings is an $N$-fold
cover of the torus. The pertinent combinatorics is summarized by the
action of the Hecke operator~\cite{DMVV1} which maps a modular form
into another one with the same weight. The action of the Hecke algebra
admits an interpretation in terms of the creation of a long string
background along with the addition of short string excitations to
it~\cite{GMMS1}. In this comparison it is of course more natural to
work with the grand canonical partition function by taking an ensemble
of sigma-models on ${\rm Sym}^N(\real^8)$ for all~$N\in\nat$.

In this paper we will examine this correspondence for the interacting
string theory with $g_s>0$ which arises by relaxing the free string
infrared limit. This is obtained by perturbing the orbifold conformal
field theory on (\ref{symprodorb}). To leading order, this
perturbation is described by the Dijkgraaf-Verlinde-Verlinde (DVV)
twist field~\cite{DVV1} which perturbs the free Hamiltonian $H$ via
the density
\beq
V_{\rm int}=g_s\,\sum_{1\leq a<b\leq N}\,\left(\tau^i\,\Sigma_i\otimes
\overline{\tau}{}^{\,j}\,\overline{\Sigma}{}_j\right)_{a,b}+
O\left(g_s^2\right) \ ,
\label{DVVop}\eeq
where $\tau^i$, $i=1,\dots,8$ are the excited bosonic twist fields and
$\Sigma_i$ are the fermionic spin fields. This defines a conformal
field of weight $(\frac32,\frac32)$ which is the unique least
irrelevant perturbation that preserves $Spin(8)$ spacetime rotations
and spacetime supersymmetry, and which creates a square-root branch cut in the
sigma-model with coordinates $x_a^i-x_b^i$. It intertwines between
different topological sectors of the worldsheet theory on
$\torus_\tau^2$ that are related by a basic splitting and joining
interaction between pairs of strings. Thus if we use the Hamiltonian
density (\ref{DVVop}) for computing scattering amplitudes via standard
perturbation theory, then we should reproduce the conventional
perturbative expansion of Type~IIA superstring theory~\cite{ArFro1}. This
expectation is supported by the fact~\cite{DijkMotl1} that the DVV
twist field exactly reproduces the Lorentz-invariant Mandelstam cubic
interaction vertex that describes the joining and splitting of Type~II
strings in light-cone gauge. Analysis of higher-order contact terms
reveals that the structure of superstring field theory simplifies when
expressed in terms of twist field
correlators~\cite{DijkMotl1,Moriyama1}.

In DLCQ string theory at finite temperature, the $g$-loop free energy
receives contributions from only those genus $g$ string worldsheets
which are branched covers of the spacetime torus
$\torus^2_\tau$~\cite{GOPS1}. This gives a partial discretization of
the moduli space ${\cal M}_g$ of genus $g$ Riemann surfaces which
reduces its complex dimension from $3g-3$ to $2g-3$ (from $1$ to $0$
for $g=1$). Thus perturbative string theory can be formulated entirely
in terms of covering Riemann surfaces, a scenario familiar from the
Gross-Taylor string expansion of the two-dimensional Yang-Mills
theory~\cite{GrossTaylor1,CMR1}. In this paper we will work
out explicitly the two-loop free energy which is computed from genus
two worldsheets which are branched covers of $\torus_\tau^2$. A
surface of genus two can be realized as a double cover of the complex
plane with three distinct branch cuts. Since any elliptic curve is a
double cover of the plane with two branch cuts, a genus two surface
can be built from two tori by identifying one of their branch cuts and
gluing them together along the cut. This means that the two-loop
partition function should coincide with the correlator of two twist
fields (\ref{DVVop}) in the symmetric orbifold conformal field theory
on $\torus_\tau^2$. Such a coincidence is not entirely surprising,
given that correlation functions of twist fields can be computed by
means of free string partition functions on the appropriate covering
space~\cite{LuninMath1}. Indeed, many aspects of string theory (at zero
temperature) can be recovered from the sigma-model with target space
(\ref{symprodorb})~\cite{ArFro1,JMR1}. However, while the twist field
correlator appears to be expressed in terms of branch point loci, the
DLCQ string free energy is naturally parametrized in terms of pinching
parameters corresponding to the sewing construction of the genus two cover from
an unramified covering of the spacetime torus and an auxilliary
torus. This suggests an interpretation of the correlation function $\langle
V_{\rm int}\,V_{\rm int}\rangle$ as the overlap between a long string
state and a fundamental string state, a result which is consistent
with the physical interpretation of the Hecke algebra mentioned
above. We will leave the detailed comparison of our results to
twist field correlators for future work. Here we perform the explicit
calculations required in DLCQ string theory to check these and other
correspondences, as well as to elucidate general higher-loop aspects of
perturbative string theory in the branched cover instanton
representation.

Our results also pertain to some other contexts. Foremost among these
is a two-loop order check of the matrix string theory
conjecture~\cite{DVV1,Motl1,BanksSeiberg1}. With
the notations set above, matrix strings at finite temperature are
described by maximally supersymmetric Yang-Mills gauge theory on
$\torus_\tau^2$ with gauge group $U(N)$~\cite{GS1}. The string
coupling constant $g_s$ is related to the Yang-Mills coupling constant
$g_{\rm YM}^{~}$ through
\beq
g_s=\frac1{g^{~}_{\rm YM}\,\ell_s} \ ,
\label{gsgYMrel}\eeq
where $\ell_s=\sqrt{\alpha'}$ is the string length. Thus at weak
string coupling the supersymmetric Yang-Mills theory becomes strongly
coupled and approaches a superconformal infrared fixed point which is
believed to be the supersymmetric orbifold sigma-model discussed
above~\cite{DVV1}. The DVV
twist field (\ref{DVVop}) is then the least irrelevant operator which
preserves the Yang-Mills supersymmetry. The string degrees of freedom
which emerge in the perturbative string limit $g_s\to0$ are
simultaneous eigenvalues of the matrices. At finite temperature the
matrices are defined on the torus $\torus^2_\tau$ and their
eigenvalues, which solve polynomial equations, are functions on
branched covers of $\torus^2_\tau$~\cite{GS1}. At leading order, the
thermodynamic free energy of matrix string theory arising from
summing over unbranched covers coincides with the DLCQ free
energy~\cite{GS1,GOPS1}. At
next order, instantons interpolate between initial and final states of
strings through the genus two cover of the
torus~\cite{Wynter1,BBN1,BBNT1,Brax1,BDAP1,GHV1}. By finding the BPS
instantons which represent the branch points~\cite{BBNT1,KoganSz1} and
the correct instanton measure, the matrix model partition function
should coincide with the twist field correlator of the symmetric
orbifold model in the strong coupling $g^{~}_{\rm YM}\to\infty$ limit
which is computed by our genus two DLCQ free energy. For this
comparison, the most appropriate DLCQ theory is that of the Green-Schwarz
superstring at finite temperature with worldsheets of long strings.

Our detailed computations and results could also shed further light on
aspects of more complicated symmetric orbifold conformal field
theories. An important example is when the orbifold target space is
taken to be ${\rm Sym}^N({\cal M})$ with ${\cal M}={\rm K}3$ or
${\cal M}=\torus^4$~\cite{AdSCFTrev}. With $N=k\,n$, a particular
deformation of the superconformal field theory is the sigma-model on
the moduli space of $k$ instantons in $U(n)$ gauge theory on $\cal M$
which is believed to be dual, via the AdS/CFT correspondence, to
Type~II string theory on the background geometry ${\rm
  AdS}_3\times\mathbb{S}^3\times{\cal M}$. The primary evidence for
these particular correspondences comes from the matching of their BPS
spectra. Finally, from a mathematical perspective our results are
related to the computation of elliptic genera~\cite{DMVV1,Dijk1} and
topological Euler characteristics of Hilbert schemes~\cite{Gotsche1},
which in the case ${\cal M}={\rm K}3$ is related to generalized
Kac-Moody algebras~\cite{Kawai1,GritNik1}.

The organisation of the remainder of this paper is as follows. In
Section~2 we review the basic arguments establishing that DLCQ string
theory at finite temperature is a theory of branched coverings of a
torus~\cite{GOPS1}. We also outline some generic aspects of a certain
reduction technique for the Hurwitz moduli space of branched covers
which will be central to our analysis throughout this paper. We
conclude by reviewing the one-loop calculation~\cite{GS1,GOPS1} in
this light for later comparison with the two-loop results.

In Section~3 we begin the construction of the two-loop free energy. We
present an explicit description of the moduli space of genus two
branched covers using a particular reduction technique. As an example,
we compute the bosonic free energy in terms of genus one
theta-functions of the elliptic curve $\torus^2_\tau$. While bosonic
string theory cannot emerge from a gauge theory (since the necessary
supersymmetric cancellations of fluctuation determinants do not
occur), this calculation can be compared to the bosonic sigma-model
with target space ${\rm Sym}^N(\real^{24})$ and the interaction
density (\ref{DVVop}) modified by replacing $\tau^i\,\Sigma_i$ with
the unexcited twist field~\cite{Rey1}
\beq
\sigma=\prod_{i=1}^{24}\,\sigma^i
\label{sigmaunexc}\eeq
of dimension $\frac32$ having the supersymmetry variation $G^{\dot
  a}_{-1/2}(\sigma\,\Sigma_{\dot a})=\tau^i\,\Sigma_i$.

In Section~4 we compute the two-loop superstring free energy. Our
calculation draws heavily on recent progress~\cite{DHPh1} in two-loop
superstring perturbation theory in the NSR formalism which yields explicit
unambiguous expressions for the chiral superstring measure in terms of
genus two modular forms. With the appropriate modification of the
genus two GSO projection at finite temperature~\cite{AtWitten1}, we
find a formula for the superstring free energy in terms of
theta-functions on $\torus_\tau^2$. In Section~5 we perform the
analogous calculation for the heterotic string. In this case the
pertinent conformal field theory is the supersymmetric heterotic
sigma-model defined on the symmetric product
orbifold~\cite{Rey1,Lowe1}
\beq
{\rm Sym}^N\bigl(\real^8\times G\bigr)=(\real^8\times G\bigr)^N/
S_N\ltimes(\zed_2)^N
\label{hetsymorb}\eeq
for the heterotic gauge group $G$. The interaction density
(\ref{DVVop}) should be modified to contain the bosonic twist field
$\overline{\sigma}$ given by (\ref{sigmaunexc}) in the right-moving
sector and the supersymmetric twist field ${\tau}{}^{i}\,{\Sigma}{}_i$
in the left-moving sector. The relevant gauge dynamics is conjectured
to be governed by heterotic matrix string
theory~\cite{Rey1,Horava1,BanksMotl1,Lowe1,BHeterotic1}, i.e. two-dimensional
supersymmetric Yang-Mills theory with chiral anomaly-free matter
fields and gauge group $O(N)$.

Our formulas for the free energies, while in principle being explicit,
are quite complicated. In Section~6 we consider various degeneration
limits of the genus two covers in which these expressions drastically
simplify, and hence elucidate various arithmetic and
physical properties of our amplitudes. We find the appropriate
modification of the action of the Hecke algebra for twist field
correlators. In a certain collapsing limit, we also find effective
one-loop string theories which resemble non-supersymmetric strings on
particular $\zed_2$-orbifolds. In another collapsing limit, the
partition function resembles the one-loop instanton sum over long string
configurations. Finally, in Appendix~A we present an alternative reductive
description of the moduli space of genus two branched covers which may
be of independent interest and use in other applications, while
Appendix~B contains some technical details of the calculations
performed in the main text.

\newsection{String Worldsheets in Light-Like Compactifications}

In this section we shall describe the general set-up for the
calculation that we will undertake. The main new technical tool we
introduce is the method of reduction which works for any branched
covering of a Riemann surface by another Riemann surface. We shall
also review the well-known one-loop calculation in this new light for
the purpose of later comparison and because it will play a role in
some of our analysis at two-loops in subsequent sections.

\subsection{Discrete Light-Cone Quantization of String Theory\\ at Finite
Temperature}

Consider the discrete light-cone quantization (DLCQ) of Type II
superstring theory at finite temperature using the Polyakov path
integral~\cite{GOPS1}. We work throughout in the Neveu-Schwarz-Ramond
formalism. In string perturbation theory, the gauge-fixed action in
the conformal gauge and in Euclidean spacetime at genus $g$ is
$S[X]+\overline{S[X]}+S[B,C]+\overline{S[B,C]}$, where
\beq
S[X]+S[B,C]=\frac1{4\pi\,\alpha'}\,\int\limits_{\Sigma_g}\dd^2z~
\Bigl(\mbox{$\frac12$}\,|\partial x^\mu|^2
+\psi_\mu\,\overline{\partial}\psi^\mu+b\,\overline{\partial}c+\beta\,
\overline{\partial}\gamma\Bigr)
\label{Polyakovaction}\eeq
and $\sqrt{\alpha'}$ is the string scale. Here
$X=(x^\mu,\psi^\mu)_{\mu=0}^9$ denotes the spacetime matter fields, while
$B$ and $C$ denote the $b,\beta$ and $c,\gamma$ ghost fields, respectively,
with $(b,c)$ the spin $(2,1)$ conformal ghost fields and $(\beta,\gamma)$ the
spin $(\frac32,\frac12)$ superconformal ghost fields. The worldsheet
is an oriented compact Riemann surface $\Sigma_g$ of genus $g$ whose first
homology group is generated by a set of canonical one-cycles $\mbf
a=(a_i)_{i=1}^g$, $\mbf b=(b_i)_{i=1}^g$ with intersection numbers
\beq
a_i\cap a_j=b_i\cap b_j=0~~~~~~,~~~~~~a_i\cap b_j=-b_j\cap a_i=\delta_{ij} \ .
\label{intnumbers}\eeq
This intersection form is summarized by the matrix
\beq
{\sf
  J}_g=\left(\begin{array}{cc}\mbf0_g&\id_g\\-\id_g&\mbf0_g\end{array}
\right)
\label{intmatrixg}\eeq
with ${\sf J}^2_g=-\id_g$ which makes $H_1(\Sigma_g,\real)$ into a
symplectic vector space.
The first cohomology group $H^{1,0}(\Sigma_g,\complex)$ is spanned by a set of
holomorphic one-differentials $\mbf\omega=(\omega_i)_{i=1}^g$ which have the
period normalizations
\beq
\oint\limits_{a_i}\omega_j=\delta_{ij}~~~~~~,~~~~~~\oint\limits_{b_i}\omega_j=
\Omega_{ij} \ ,
\label{periodnorms}\eeq
where $\Omega$ is the period matrix of $\Sigma_g$ which lives in the Siegal
upper half-plane ${\cal H}_g$ of $g\times g$ complex-valued, symmetric matrices
of positive definite imaginary part. We shall throughout write
$\Omega=\Omega_1+\ii\Omega_2$, where $\Omega_1$ and $\Omega_2$ are real-valued
symmetric matrices with $\Omega_2>0$.

The DLCQ and finite temperature conditions are imposed by two spacetime
compactifications which may be described by the respective identifications
\bea
\left(x^0\,,\,\mbf x\,,\,x^9\right)&\sim&\big(x^0+\sqrt2\,\pi\ii R\,,\,\mbf x
\,,\,x^9-\sqrt2\,\pi\,R\big) \ , \nn\\
\left(x^0\,,\,\mbf x\,,\,x^9\right)&\sim&\left(x^0+\beta\,,\,\mbf
  x\,,\,x^9\right)
\label{compconds}\eea
where $R$ is the radius of the light-cone in Minkowski space, and
$\beta=1/k_{\rm B}\,T$ with $T$ the temperature and $k_{\rm B}$ the Boltzmann
constant. The corresponding path integral, with the appropriate modification of
the GSO projection to make spacetime fermions anti-periodic under $x^0\to
x^0+\beta$, then computes the thermodynamic free energy of the superstring. The
compactification conditions induce quantized zero modes in the mode expansions
of the bosonic string embedding fields $x^\mu$ corresponding to the wrappings
of the various homology cycles of $\Sigma_g$ around the compact spacetime
dimensions. The windings of $(\mbf a,\mbf b)$ around the
light-cone are labelled by integers $(\mbf p,\mbf q)$ and by $(\mbf n,\mbf
m)$ around the time direction. Apart from the modification of the GSO
projection by the temperature winding numbers $(\mbf n,\mbf m)$, the only
place that these integers appear are as zero mode soliton contributions to the
bosonic matter part of the action (\ref{Polyakovaction}). In the path integral
one should sum over all possible topological winding sectors. The crucial point
is that the action (\ref{Polyakovaction}) depends linearly in a purely
imaginary form on the set of integers $(\mbf p,\mbf q)$, which when summed
thereby produce periodic Dirac delta-functions.

In this way, the finite-temperature, DLCQ superstring free energy (per
unit spacetime volume) at genus $g$ is found to be given
by~\cite{GOPS1}
\bea
F_g&=&-g_s^{2g-2}\,\nu^{2g}\,\sum_{\mbf m,\mbf n\in\zeds^g}~
\sum_{\mbf r,\mbf s\in\zeds^g}~~\int\limits_{{\cal F}_g}\dd\mu_g
\left[{\mbf n}\atop{\mbf m}\right]\bigl(\Omega\,,\,\overline{\Omega}\,
\bigr)~\det\Omega_2~\e^{-\frac{\beta^2}{4\pi\,\alpha'}\,
(\Omega\mbf n-\mbf m)^\dagger\,(\Omega_2)^{-1}(\Omega\mbf n-\mbf m)}\nn\\&
&\times\,\prod_{j=1}^g\delta\left(\,\mbox{$\sum\limits_{i=1}^g$}\,(n_i+\ii\nu
\,r_i)\,\Omega_{ij}-(m_j+\ii\nu\,s_j)\right) \ ,
\label{calZg}\eea
where $g_s$ is the string coupling constant and
\beq
\nu=\frac{4\pi\,\alpha'}{\sqrt2\,\beta\,R} \ .
\label{nudef}\eeq
The sums in (\ref{calZg}) go over all four $g$-vectors of integers $\mbf m,\mbf
n,\mbf r,\mbf s$ such that the period matrix $\Omega$ is in a fundamental
modular domain ${\cal F}_g$. The modular invariant, genus $g$ superstring
measure on moduli space ${\cal M}_g$ is denoted $\dd\mu_g[{{\mbf n}\atop{\mbf
m}}](\Omega,\overline{\Omega}\,)$, and its dependence on the
temperature winding integers arises
from the modification of the sum over worldsheet spin structures that breaks
supersymmetry in the finite temperature theory \cite{AtWitten1}. The
expression (\ref{calZg}) contains a constraint on the Riemann surfaces
$\Sigma_g$ which contribute to the partition function. As we now
explain, it is equivalent to summing over all genus $g$ branched
covers $\Sigma_g$ of the torus $\torus^2_{\ii\nu}$ whose Teichm\"uller
parameter is $\ii\nu$~\cite{GOPS1}.

Let $f:\Sigma_g\rightarrow\torus^2_{\ii\nu}$ be a holomorphic map,
i.e. a branched covering. The covering map induces a homomorphism
between the first homology groups via the push-forward
\beq
f_*\,:\,H_1(\Sigma_g,\zed)~\longrightarrow~H_1(\torus^2_{\ii\nu},
\zed) \ .
\label{pushforward}\eeq
Choosing canonical homology bases $(\mbf a,\mbf b)$ and
$(\alpha,\beta)$ of the covering space $\Sigma_g$ and the base space
$\torus^2_{\ii\nu}$, respectively, this homomorphism can be written
explicitly in terms of an integral $2\times2g$ matrix
\beq
\homo=\left(\begin{array}{cc}\mbf n&\mbf m\\\mbf r&\mbf s\end{array}
\right)
\label{hommatrixints}\eeq
of maximal rank acting on the homology generators of the base torus as
\beq
f_*\left(\begin{array}{c}\mbf a\\\mbf b\end{array}\right)=\homo^\top
\left(\begin{array}{c}\alpha\\\beta\end{array}\right) \ .
\label{homologycover}\eeq
Similarly, the covering map induces through pull-back a homomorphism
$f^*:H^{1,0}(\torus^2_{\ii\nu},\complex)\to H^{1,0}(\Sigma_g,\complex)$ on the
first
cohomology groups, and there exists a complex $g\times1$ matrix
$\pull$ of maximal rank which relates the normalized holomorphic differentials
$\mbf\omega$ and $\omega$ on $\Sigma_g$ and $\torus^2_{\ii\nu}$
by
\beq
\omega=\pull^\top\mbf\omega \ .
\label{pulldef}\eeq
The matrix $\pull$ can be used to give an explicit formula for the
covering map as
$f(z)=(\Psi\circ\mbf{\mathfrak{A}})(z):=\pull^\top\mbf{\mathfrak{A}}(z)~{\rm
  mod}~\zed\oplus\ii\nu\,\zed$, where $\mbf{\mathfrak{A}}$ is the Abel
map embedding $\Sigma_g$ into its Jacobian variety ${\rm
  Jac}(\Sigma_g):=\complex^g/\zed^g\oplus\Omega\,\zed^g$. This
characterization exploits the fact that the Jacobian variety of the
curve $\Sigma_g$ represents a fibration over the elliptic curve
$\torus_{\ii\nu}^2$ with the commutative diagram
\beq
\xymatrix{ \Sigma_g \ar[r]^{\!\!\!\!\!\!\!\!\!\mbf{\mathfrak{A}}} \ar[rd]_f&
{\rm Jac}(\Sigma_g) \ . \ar[d]^\Psi\\ & \torus_{\ii\nu}^2 }
\label{Abelcommdiag}\eeq
Furthermore, by computing the $\alpha$ and $\beta$ periods of both
sides of (\ref{pulldef}) we arrive at the matrix equality
\beq
\pull^\top\,(\id_g,\Omega)=(1,\ii\nu)\,\homo \ .
\label{matrixpulleq}\eeq
By using the explicit form (\ref{hommatrixints}) one finds $\pull=\mbf
n+\ii\nu\,\mbf r$ and the equation (\ref{matrixpulleq}) is equivalent
to the period matrix constraint in (\ref{calZg}).

The degree ${\rm deg}(f)$ of the covering map can be computed from the
Hopf condition~\cite{Martens1}
\beq
\homo\,\sympl_g\,\homo^\top={\rm deg}(f)~\sympl_1
\label{Hopfcond}\eeq
giving ${\rm deg}(f)=\mbf n\cdot\mbf s-\mbf m\cdot\mbf r$. The
computation of the periods in (\ref{pulldef}) leads to a homogeneous
linear equation in the variables $\mbf n,\mbf m,\mbf r,\mbf s$ and
$\ii\nu$ which has the compatibility condition
\beq
\det\left(\begin{array}{ccc}\id_g& &\Omega\\\dots&\dots&\dots\\ &\homo&
\end{array}\right)=0 \ .
\label{compcondperiods}\eeq
The formula (\ref{compcondperiods}) restricts the allowed Riemann
period matrices of the curve $\Sigma_g$ to lie in a Humbert variety
inside $\mathcal{H}_g$.

\subsection{Weierstrass-Poincar\'e Reduction\label{W-PRed}}

The superstring integration measure $\dd\mu_g[{{\mbf n}\atop{\mbf
m}}](\Omega,\overline{\Omega}\,)$ is invariant
under the mapping class group of the covering Riemann surface. This
invariance can be exploited in a manner which simplifies explicit
calculations. The Siegal modular group of $\Sigma_g$ is $Sp(2g,\zed)$
which preserves the intersection form (\ref{intmatrixg}). With respect
to the canonical basis of $\real^{g,g}$, it consists of matrices
\beq
\left(\begin{array}{cc}A&B\\ C&D\end{array}\right)~~~~,~~~~D^\top
\,B-B^\top\,D=C^\top\,A-A^\top\,C=0
{}~~~~,~~~~A^\top\,D-C^\top\,B=\id_g \ .
\label{modulargroup}\eeq
This group acts on a canonical homology basis of $\Sigma_g$ as
\beq
\left(\begin{array}{c}\mbf{a}\\ \mbf{b}\end{array}\right)~\longmapsto~
\left(\begin{array}{c}\mbf a^{\,\prime}\\
\mbf b^{\,\prime}\end{array}\right)=\left(\begin{array}{cc}D&C\\
B&A\end{array}\right)\left(\begin{array}{c}\mbf{a}\\
\mbf{b}\end{array}\right) \ .
\label{homologymodular}\eeq
The temperature winding integers $(\mbf n,\mbf m)$ transform in the same way as
$(\mbf a,\mbf b)$, and so do the integers $(\mbf r,\mbf s)$ which come from
compactification of the light-cone. Using (\ref{modulargroup}) the inverse of
the transformation (\ref{homologymodular}) is easily found to be
\beq
\left(\begin{array}{c}\mbf{a}\\ \mbf{b}\end{array}\right)=
\left(\begin{array}{cc}A^{\top}&-C^\top\\
-B^\top&D^\top\end{array}\right)\left(\begin{array}{c}\mbf a^{\,\prime}\\
\mbf b^{\,\prime}\end{array}\right) \ .
\label{modularinv}\eeq
The projective modular group $PSp(2g,\zed)$ acts naturally on the Siegal upper
half-plane ${\cal H}_g$ of $g\times g$ period matrices as
\beq
\Omega~\longmapsto~\Omega'=(A\,\Omega+B)\,(C\,\Omega+D)^{-1} \ .
\label{Omegamodular}\eeq
For genera $g=1,2,3$, the moduli space of $\Sigma_g$ is ``almost''
given by~\cite{Morozov1}
\beq
{\cal M}_g={\cal H}_g/PSp(2g,\zed) \ .
\label{calMg}\eeq
For $g\geq4$ the period matrix can still be used to parametrize moduli
space by imposing Schottky relations on $\Omega$. Note that the
delta-function constraint of (\ref{calZg}) is modular covariant.

We can now use the technique of reduction to simplify the constraint
equation on $\Omega$ in (\ref{calZg}) before solving it. The technique
of reduction is described in~\cite{Martens1} (see
also~\cite{Igusa1,BelEn1}) for the general case of coverings of Riemann
surfaces of arbitrary genus. Reduction comprises the use of modular
transformations on the base and cover in order to make a
change in homology basis so that the number of homology cycles on the
cover which effectively wind around the base is reduced. It yields a convenient
canonical form for the underlying algebraic curve $\Sigma_g$ which can
be thought of as a higher genus version of the canonical Weierstrass
parametrization of an elliptic curve. In the present case the periods
meet the conditions of the fundamental Weierstrass-Poincar\'e theory
of the complete reducibility of abelian integrals to lower
genera~\cite{Martens1}, which deals with general abelian tori and
their associated theta-functions. The main idea is that the curve
$\Sigma_g$, being a covering of a torus, has a rich group of
automorphisms which leads to a decomposition of its Jacobian
variety. By considering the curve as a spectral variety, one can
thereby reduce the corresponding theta-functions to lower
genera. Furthermore, the technique greatly simplifies the analysis of
moduli space integrals such as (\ref{calZg}) by extending the usual
Rankin-Selberg method of ``unwrapping'' modular
integrals~\cite{McClainRoth1}.

We will use the Poincar\'e reducibility theorem applied to the special
case of a covering $f:\Sigma_g\to\torus^2_{\ii\nu}$. It
relies~\cite{Martens1} on the existence of a Frobenius normal form for
the $2\times2g$ integral matrix (\ref{hommatrixints}), satisfying the
Hopf condition (\ref{Hopfcond}), given by
\beq
\homo={\sf S}\,{\sf P}\,{\sf T}
\label{Frobform}\eeq
where $\sf S$ and $\sf T$ are, respectively, $2\times2$ and
$2g\times2g$ symplectic unimodular matrices. The Poincar\'e normal
form is given by the $2\times2g$ matrix
\be \label{PNF}
{\sf P}=r\,\left(\begin{array}{ccccccccccc}
1 & 0 & 0 &\ldots & 0 & & 0 & 0 & 0 &
\ldots ~~ 0 \\ 0 & s & 0 &\ldots & 0 & & t & 0 & 0 &\ldots ~~ 0
\end{array}\right) \ ,
\ee
where $r,s,t$ are integers such that $r^2\,t={\rm deg}(f)$ and $s$
either vanishes or is a divisor of $t$. The cases $s=0$ can be
ruled out by the requirement that block diagonal period
matrices are not allowed~\cite{Moore1}, being contributions from a
particular boundary component of moduli space. The existence of the
form (\ref{PNF}) implies, among other things, that there exists a
modular transformation such that the windings around the temperature
direction of spacetime occur {\it only} around the single homology
cycle $a_1$, with all other cycles being periodic. This means that the
compactification conditions can be chosen to be
\bea
\oint\limits_{a_1'}\,\dd x^0&=&\beta\,r+\sqrt2\,\pi\,R\ii p_1' \ ,
\nonumber\\
\oint\limits_{a_j'}\,\dd x^0&=&\sqrt2\,\pi\,R\ii p_j' \ , \quad
j=2,\dots,g \ , \nonumber\\
\oint\limits_{b_i'}\,\dd x^0&=&\sqrt2\,\pi\,R\ii q_i' \ ,
\nonumber\\
\oint\limits_{a_i'}\,\dd x^9&=&\sqrt2\,\pi\,R\,p_i' \ ,
\nonumber\\
\oint\limits_{b_i'}\,\dd x^9&=&\sqrt2\,\pi\,R\,q_i'
\label{compcondsPNF}\eea
for $i=1,\dots,g$, with the transverse components $\mbf x$ periodic
around the new basis of homology cycles $\mbf a',\mbf b'$ of
$\Sigma_g$. In addition, after summation over $\mbf p',\mbf q'$ only
the homology cycles $a_2$ and $b_1$ wrap around the light-cone.

Reduction depends on the number theoretic properties of the entries of
the integral matrix $\sf M$ and is explicitly carried out by using the
$2g\times2g$ symplectic matrices
\beq
\left(\begin{array}{cc}\id_g& S\\ \mbf 0_g&
    \id_g\end{array}\right)~~~~,~~~~\left(\begin{array}{cc}\id_g&
\mbf 0_g\\ S& \id_g\end{array}\right)~~~~,~~~~
\left(\begin{array}{cc}A& \mbf0_g\\ \mbf0_g& \left(A^{-1}\right)^\top
\end{array}\right) \ ,
\label{redmatrices}\eeq
where $S$ is a symmetric $g\times g$ integral matrix and $A\in
SL(g,\zed)$. By regarding the matrix (\ref{hommatrixints}) as
consisting of two $2\times g$ block matrices
$\homo=(\homo_1,\homo_2)$, the matrices (\ref{redmatrices})
interpolate between these blocks via elementary row and column
operations. Using the normal form (\ref{Frobform})
one can transform (\ref{matrixpulleq}) into the equation of the
Weierstrass-Poincar\'e theorem
\beq
(\id_g,\Omega)\,{\sf T}={\sf F}\,\left(\begin{array}{cccccc}1&0&\dots&0&
-\frac{\sigma_1}{t\,\sigma_2}&\mbf q\\ \mbf 0_{(g-1)\times1}&
&\id_{g-1}&\mbf q^\top&{\sf Z}\end{array}\right) \ ,
\label{WPthm}\eeq
where $\sf F$ is a non-singular $g\times g$ complex matrix, $\mbf
q=(-\frac st,0,\dots,0)$ is a $(g-1)$-vector, the complex numbers
$\sigma_1,\sigma_2$ are defined by
$(\sigma_1,\sigma_2)=(1,\ii\nu)\,{\sf S}$, and $\sf Z$ is a
$(g-1)\times(g-1)$ complex matrix satisfying the Riemann bilinear
relations which can be found after explicit construction of the
symplectic transformation. Because the vector $\mbf q$ is
rational-valued, the corresponding genus~$g$ theta-functions factorize
into products of theta-functions of lower genera based on the curves
with periods $\frac{\sigma_1}{t\,\sigma_2}$ and $\sf Z$.

The reduction to the Poincar\'e normal form (\ref{PNF}) can
be thought of as a gauge fixing of the large diffeomorphism symmetry (the
mapping class group) of the Riemann surface $\Sigma_g$. There is still
then a residual gauge symmetry left over, which we will fix by restricting to
those modular transformations which preserve the corresponding reduced
compactification conditions. This defines a proper subgroup ${\cal
G}\subset Sp(2g,\zed)$, and so it will {\it extend} the fundamental modular
region for the action of $Sp(2g,\zed)$ on ${\cal H}_g$ from ${\cal F}_g$ to
some domain ${\cal F}_g'$. Modular invariance is then restored via the
observation \cite{McClainRoth1} that the new region ${\cal F}_g'$ is composed
of an
infinite number of images of the fundamental domain ${\cal F}_g$ under certain
elements of the modular group. The sum over all copies of ${\cal F}_g$ in
${\cal F}_g'$ may be implemented by a sum over all elements of the coset
$Sp(2g,\zed)/{\cal G}$. The corresponding constraints on the period
matrix $\Omega$ in (\ref{WPthm}) reduce the complex dimension $3g-3$
of moduli space to $2g-3$. In addition, there is discrete data
contained in the compactification integers, such as those
arising from the requirement that the real-valued symmetric matrix
$\Omega_2$ be positive. This gives a partial discretization of the
Riemann moduli space ${\cal M}_g$ to the Hurwitz moduli space of
holomorphic maps, with the $2g-3$ moduli given by the branching data
required to build the cover $\Sigma_g$ from its base
$\torus^2_{\ii\nu}$. The Hurwitz space can be embedded as an analytic
subvariety of ${\cal M}_g$~\cite{FarkasKra1,Forster1}.

\subsection{One-Loop Computation\label{OneLoop}}

It is instructive to recall the genus one
situation~\cite{GS1,GOPS1}. Then all covers
$\Sigma_1\to\torus^2_{\ii\nu}$ are unbranched. In this case one can
deduce the period constraint of (\ref{calZg}) by elementary methods
which exhibit the geometric construction of the covering torus from
the base torus in terms of the compactification integers specified by
(\ref{hommatrixints}). For this, let us regard the torus
$\torus_{\ii\nu}^2$ as the quotient of the complex plane $\complex$ by
a lattice $\Lambda=\left<e_1,e_2\right>:=\zed\,e_1\oplus\zed\,e_2$ of
rank~$2$ generated by two-vectors $e_1$ and $e_2$. The isomorphism
classes of unramified covers $\Sigma_1\to\torus_{\ii\nu}^2$ of degree
$N$ then correspond to the inequivalent sublattices $\Lambda'\subset
\Lambda$ of index $[\Lambda:\Lambda'\,]=N$. These may be found as
follows. Let $f_1=r'\,e_1\in\Lambda'$ be the smallest multiple of
$e_1$. Then there exists $f_2=s'\,e_1+m'\,e_2\in \Lambda'$ with
$s'<r'$ such that $\Lambda'$ is generated by $f_1$ and $f_2$ over
$\zed$. The index of this lattice is $r'\,m'$. As a consequence, for
each integer $r'$ dividing the index $N$ there are $r'$ inequivalent
sub-lattices
\be\label{possiblesublattices}
\left<r'\,e_1,\mbox{$\frac N{r'}$}\,e_2\right>~~,~~
\left<r'\,e_1,e_1+\mbox{$\frac N{r'}$}\,e_2\right>~~,~~\ldots~~,~~
\left<r'\,e_1,(r'-1)\,e_1+\mbox{$\frac N{r'}$}\,e_2\right> \ .
\ee
It follows that the number of inequivalent sublattices
$\Lambda'\subset \Lambda$ of index $[\Lambda : \Lambda'\,]=N$ is
\begin{equation}
\sigma_1(N)=\sum_{r'|N}\,r' \ ,
\label{sigma1Ndef}\end{equation}
and the moduli of the corresponding covers are given by
\beq
\tau=\frac{s'+\frac\ii\nu\,m'}{r'} \ .
\label{tauexpl}\eeq

We will now use the Weierstrass-Poincar\'e reduction to show that
solving the reduced constraint in this case gives the same moduli
(\ref{tauexpl}) of the covers constructed from the base modular parameter
$\ii\nu$. The integers $n'=0,m'\in\zed$ are
defined by the $SL(2,\zed)$ transformation
\beq
n'=0=D\,n+C\,m~~~~~~,~~~~~~-m'=B\,n+A\,m \ .
\label{genus1modular}\eeq
The first equation is solved by the relatively prime integers $C=-n/{\rm
gcd}(n,m)$ and $D=m/{\rm gcd}(n,m)$. Now we use the fact that the set of
integers $\zed$ is a principal ideal domain, which implies that there exists
integers $A$ and $B$ such that
\beq
A\,m+B\,n={\rm gcd}(n,m) \ .
\label{AD-BC=1}\eeq
Reduction for the genus one case is thus very simple, as all the
windings of the cover $\Sigma_1$ around the temperature direction are
put into the $b$ cycle by the $SL(2,\zed)$ transformation generated by
the unimodular matrix
\beq
{\sf T}_1=\left(\begin{array}{cc}\frac m{{\rm gcd}(n,m)}&B\\
-\frac n{{\rm gcd}(n,m)}&A\end{array}\right) \ .
\label{Xi1}\eeq
Furthermore, from (\ref{genus1modular}) it follows that the sole
temperature winding integer is given by the greatest common divisor of
the original two winding numbers as
\beq
m'=-{\rm gcd}(n,m) \ .
\label{mprime}\eeq

The constraint equation for the modulus $\tau$ of the cover $\Sigma_1$
is given by
\beq
\pull^\top\,(1,\tau)=(1,\ii\nu)\,\left(\begin{array}{cc}n&m\\r&s\end{array}
\right)=(1,\ii\nu)\,\left(\begin{array}{cc}0&-m'\\r'&s'\end{array}
\right)\,\left(\begin{array}{cc}A&-B\\\frac n{{\rm gcd}(n,m)}&
\frac m{{\rm gcd}(n,m)}\end{array}\right) \ ,
\label{genus1constreq}\eeq
which can be solved explicitly to determine $\tau$ as in
(\ref{tauexpl}) with
\beq
r'=\frac{m\,r-n\,s}{{\rm gcd}(n,m)}~~~~~~,~~~~~~s'=B\,r+A\,s \ .
\label{newrnews1}\eeq
The genus one fundamental domain is given by
\beq
\Delta:=
{\cal F}_1=\left\{\tau\in\complex~\Bigm|~\mbox{$-\frac12<\tau_1\leq\frac12$}
{}~,~|\tau|^2\geq1~,~\tau_2>0\right\} \ .
\label{calF1}\eeq
Requiring the reduced compactification constraints to be modular
invariant sets $C=0$ and $A=D=1$ in (\ref{modulargroup}). Thus only
the translations $\tau\mapsto\tau+B$, $B\in\zed$ survive under the
action of the restricted modular group $\cal G$ on Teichm\"uller
space, and the fundamental modular region is extended to the strip
\beq
\Delta':=
{\cal F}_1'=\left\{\tau\in\complex~\Bigm|~\mbox{$-\frac12<\tau_1\leq\frac12$}
{}~,~\tau_2>0\right\} \ .
\label{calF1prime}\eeq
Requiring that $\tau\in\Delta'$ is then equivalent to $s'\in\zed/{r'}\,\zed$,
$N:=m'\,r'>0$.

The integration measure on moduli space is obtained by computing the standard
zero temperature, chiral Laplacian determinants on the torus induced by
integrating out the ten worldsheet bosonic fields $x^\mu$, their superpartners
$\psi^\mu$, and the ghosts, in a given spin structure. The GSO projection then
dictates to sum over the three even spin structures in each of the left and
right moving sectors of the worldsheet field theory (The odd spin structure
$(1,1)$ yields a vanishing contribution due to the zero modes of
the free worldsheet fermion fields $\psi^\mu$). The appropriate modification
which makes the spacetime fermion fields antiperiodic inserts an extra phase
factor $(-1)^{m'}$ in front of the GSO phase associated with the $(0,1)$
spin structure. The modular invariant, finite-temperature superstring measure
is thereby given as~\cite{AtWitten1}
\beq
\dd\mu_1^{(m')}\bigl(\tau\,,\,\overline{\tau}\,\bigr)
=\left(\frac1{4\pi^2\,\alpha'}\right)^5~
\frac{\dd^2\tau}{(\tau_2)^6}~\frac1{4\big|\eta(\tau)\big|^{8}}\,\Bigl|
\theta_2(0|\tau)^4-\theta_3(0|\tau)^4+\e^{\pi\ii m'}\,
\theta_4(0|\tau)^4\Bigr|^2 \ .
\label{dmu1}\eeq
Here the Jacobi-Erderlyi functions $\theta_a(z|\tau)$, $a=2,3,4$ (which are
induced by the spacetime fermion fields and the superconformal ghost fields)
are defined in terms of the three even characteristic, genus one
theta-functions as $\theta_2=\theta{1\choose0}$,
$\theta_3=\theta{0\choose0}$, and $\theta_4=\theta{0\choose1}$, where
\beq
\theta{a\choose b}(z|\tau)=\sum_{n=-\infty}^\infty\e^{\pi\ii
(n+\frac12\,a)^2\,\tau}~
\e^{2\pi\ii(n+\frac12\,a)\,(z+\frac12\,b)}
\label{thetagenus1}\eeq
are holomorphic functions of $(z|\tau)\in\complex\times{\cal H}_1$ for
$a,b\in\real$, while
\beq
\eta(\tau)=\mbox{$\frac12$}\,\theta_2(0|\tau)\,\theta_3(0|\tau)\,
\theta_4(0|\tau)
\label{Dedekind}\eeq
is the Dedekind function (which is induced by the spacetime boson fields and
the conformal ghost fields). By using the Jacobi abstruse identity
\beq
\theta_3(0|\tau)^4-\theta_4(0|\tau)^4-\theta_2(0|\tau)^4=0 \ ,
\label{abstruse}\eeq
we can simplify the expression (\ref{dmu1}) to
\beq
\dd\mu_1^{(m')}\bigl(\tau\,,\,\overline{\tau}\,
\bigr)=\left(\frac1{4\pi^2\,\alpha'}\right)^5~
\frac{\dd^2\tau}{(\tau_2)^6}~\frac{\left(1-\e^{\pi\ii m'}\right)
\bigl|\theta_4(0|\tau)\bigr|^8}{2\big|\eta(\tau)\big|^{8}} \ .
\label{dmu1abstruse}\eeq

By substituting all of these expressions back into the genus one free
energy (\ref{calZg}) and integrating the delta-function with the
appropriate Jacobian factor, we arrive finally at
\beq
F_1=-\frac1{\sqrt2\,\pi\,R\,\beta}\,\sum_{N=1}^\infty\,
\e^{-\frac{\beta\,N}{\sqrt2\,R}}~{\bf H}_N
*\left[\left(\frac1{4\pi^2\,\alpha'\,\tau_2}\right)^4\,
\frac{\big|\theta_4(0|\tau)\big|^8}
{\big|\eta(\tau)\big|^{8}}\right]\biggm|_{\tau=\ii/\nu} \ ,
\label{calZ1mod}\eeq
where ${\bf H}_N$ are the (restricted) Hecke operators~\cite{Apostol1}
whose actions on a modular invariant function
$f(\tau,\overline\tau\,)$ on Teichm\"uller space are defined by
\beq
{\bf H}_N*f(\tau,\overline\tau\,)=\frac1N~
\sum_{\substack{m'\,r'=N\\ m'\,\textrm{odd}}}~
\sum_{s'\in\zed/{r'}\,\zed}\,
f\left(\mbox{$\frac{s'+\tau\,m'}{r'},\frac{s'+\overline{\tau}\,m'}{r'}$}
\right) \ .
\label{Heckeop}\eeq
By applying the modular transformation $\tau\mapsto-1/\tau$ and using the
transformation rules
\beq
\theta_4(\mbox{$\frac z\tau|-\frac1\tau$})=\sqrt{-\ii\tau}~\e^{\pi\ii
z^2/\tau}~\theta_2(z|\tau)~~~~~~,~~~~~~\eta(-\mbox{$\frac1\tau$})=
\sqrt{-\ii\tau}~\eta(\tau) \ ,
\label{modtransfs1}\eeq
we can transform the expression (\ref{calZ1mod}) into the equivalent form
\beq
F_1=-\frac1{\sqrt2\,\pi\,R\,\beta}\,\sum_{N=1}^\infty\,
\e^{-\frac{\beta\,N}{\sqrt2\,R}}~{\bf H}_N
*\left[\left(\frac1{4\pi^2\,\alpha'\,\tau_2}\right)^4\,\frac{
\big|\theta_2(0|\tau)\big|^8}
{\big|\eta(\tau)\big|^{8}}\right]\biggm|_{\tau=\ii\nu} \ .
\label{calZ1}\eeq

The operand of the Hecke operators in (\ref{calZ1}) is the partition
function of a first quantized Green-Schwarz superstring, so that the
expression (\ref{calZ1}) has a natural interpretation as a map from a
first quantized to a second quantized superstring
theory~\cite{DMVV1}. The discrete Teichm\"uller parameters
(\ref{tauexpl}) indicate how the homology cycles of the base
$\torus^2_{\ii\nu}$ wind around the cycles of the unbranched cover
$\Sigma_1$. The combinatorics of enumerating unbranched covers of the
torus $\torus_{\ii\nu}^2$ are thereby elegantly
accounted for by the Hecke operators acting on the partition function of a
superconformal field theory, with toroidal worldsheet and target space
$\real^8$, in (\ref{calZ1}). This result agrees with both the
computation using operator quantization in light-cone gauge and in
matrix string theory~\cite{GS1}, as well as in the superconformal
field theory on the symmetric product orbifold
(\ref{symprodorb})~\cite{FM1}. The calculation can also be applied to
bosonic and heterotic strings, with the final result always being the
insertion of the appropriate one-loop light-cone Green-Schwarz string
partition function in the operand of the Hecke operator in
(\ref{calZ1}). In what follows we shall extend these
one-loop calculations to the case of genus two branched covers
$\Sigma_2$ of the torus $\torus^2_{\ii\nu}$.

\newsection{Bosonic Strings}

We will now extend the calculation of Section~\ref{OneLoop} by
computing the two-loop free energy $F_2$ in (\ref{calZg}). As a
warm up, in this section we will look at the simpler setting of
bosonic string theory (whose action is obtained from
(\ref{Polyakovaction}) by dropping all Grassmann fields in
26~spacetime dimensions) for which the moduli space integration measure
is more manageable. This will make the various reduction techniques
that we present more transparent. They will also carry through to the
superstring and heterotic string cases which will be studied in the
next two sections. There is a fairly complete picture of
Teichm\"uller space and moduli space at genus two. Every genus two
surface admits a hyperelliptic representation as a double cover of the
complex plane with three quadratic branch cuts supported by six branch
points. While this description is useful for describing interacting
matrix strings~\cite{Wynter1,BBNT1}, it is not the natural
parametrization for DLCQ strings.

\subsection{Two-Loop Worldsheet Contributions\label{2LoopWorld}}

The two-loop free energy is given by a sum over (equivalence classes
of) non-constant holomorphic maps
$f:\Sigma_2\to\torus_{\ii\nu}^2$. Let us begin by summarizing some
useful facts about these contributing worldsheets~\cite{CMR1}. By the
Riemann-Hurwitz theorem, the total branching number $B$ for the
branched cover of a torus by a genus two surface $\Sigma_2$ is
$B=2$. This means that a covering $f:\Sigma_2\to\torus^2_{\ii\nu}$ has
three possible types of singularities: (a) Two simple branch points;
(b) one branch point with two ramification points each of ramification
index~$2$; or (c) one branch point with one ramification point of
ramification index~$3$. The singularity types (b) and (c) can each be
thought of as degenerate limits of type~(a), which in this sense
represents the generic situation.

The lifting of curves from $\torus^2_{\ii\nu}$ to the covering space
$\Sigma_2$ induces a group homomorphism
\beq
f_\#\,:\,\pi_1\big(\torus_{\ii\nu}^2\setminus{\cal B}_f\big)~
\longrightarrow~S_N
\label{fcurvelifts}\eeq
where ${\cal B}_f\subset\torus_{\ii\nu}^2$ is the branch locus of the
covering map $f$, $N={\rm deg}(f)$, and
$\pi_1(\torus_{\ii\nu}^2\setminus{\cal
  B}_f)\cong\langle\alpha,\beta,\gamma_1,\gamma_2~|~\alpha\,\beta\,
\alpha^{-1}\,\beta^{-1}\,\gamma_1\,\gamma_2=\id\rangle$ (with
$\gamma_2=\id$ in the case that ${\cal B}_f$ consists of a single
non-simple branch point). Let
$\gamma_t$ be a homotopy generator which surrounds a branch point
$t\in{\cal B}_f\subset\torus_{\ii\nu}^2$. If $t$ is simple, then the
permutation $f_\#(\gamma_t)\in S_N$ has a single non-trivial cycle of
length~$2$. Otherwise, $f_\#(\gamma_t)$ either contains two
non-trivial cycles of length~$2$ or it has a single non-trivial cycle
of length~$3$. Together with the canonical homology generators
$\alpha,\beta$, these permutations generate a transitive subgroup
${\cal T}_{N,{\cal B}_f}$ of $S_N$ and the induced map
(\ref{fcurvelifts}) is an isomorphism onto this subgroup. There is a
one-to-one correspondence between elements of ${\cal T}_{N,{\cal
    B}_f}$ and irreducible branched covers. The two-loop free energy
that we obtain in this and the subsequent sections are thus generating
functions for the orbits in ${\cal T}_{N,{\cal B}_f}$ under
conjugation by permutations in $S_N$.

\subsection{Modular Parameters\label{Genus2Constr}}

We will now find the most general form of the $2\times2$ period
matrix $\Omega$ of the covering surface $\Sigma_2$. This will be
achieved by using a modified version of the reduction technique
described in Section~\ref{W-PRed} to solve the genus two constraint
which gives the moduli of the genus two branched
covers of the torus $\torus^2_{\ii\nu}$. The constraint equation
(\ref{matrixpulleq}) in this case reads
\be \label{matrixconstraint}\pull^\top\,(\id_2,\Omega)=
(1,\ii\nu)\,\left(\begin{array}{cccc}n_1&n_2 &m_1&m_2
    \\ r_1 &r_2& s_1&s_2\end{array}\right) \ , \ee
where $\sum_{i=1,2}\,(n_i\,s_i-m_i\,r_i)={\rm deg}(f)=:N$ is the degree
of the cover.

As in the one-loop calculation, it is possible to
calculate part of the matrix $\sf T$ appearing in the Frobenius normal
form (\ref{Frobform}) by choosing integers $A_i,B_i$ such that
\beq
A_i\,m_i+B_i\,n_i={\rm gcd}(n_i,m_i)=:n_i' \ , \quad i=1,2 \ .
\label{AiBidef}\eeq
Then the $Sp(4,\zed)$ matrix
\be\Lambda_a=\left(\begin{array}{cccc}B_1 & 0 & -\frac{m_1}{n_1'} &  0
    \\ 0 & B_2 & 0
    & -\frac{m_2}{n_2'} \\ A_1 & 0 & \frac{n_1}{n_1'} & 0 \\ 0 & A_2 & 0 &
    \frac{n_2}{ n_2'}\end{array}\right)\ee
transfers all windings from the $b_i$ homology cycles to the $a_i$
cycles, i.e. it defines a Rankin-Selberg modular transformation
(\ref{homologymodular}) for which the $2\times4$ integral matrix
$\homo$ becomes
\beq
\homo~\longmapsto~\left(\begin{array}{cccc}n_1' & n_2'& 0 & 0 \\ r_1'&
    r_2'& s_1'& s_2'\end{array}\right)
\eeq
with $r_i'=B_i\,r_i+A_i\,s_i$ and
$s_i'=\frac1{n_i'}\,(n_i\,s_i-m_i\,s_i)$ for $i=1,2$. The matrix
$\Lambda_a$ belongs to an $SL(2,\zed)\times SL(2,\zed)$
subgroup of the full modular group $PSp(4,\zed)\cong SO(3,2,\zed)$.

The next step is to move the
temperature windings from the $a_2$ cycle to the $a_1$ cycle. For this,
we introduce two further integers $U_1$ and $U_2$ with
\beq
U_1\,n_1'+U_2\,n_2'={\rm gcd}(n_1',n_2')=:r \ .
\eeq
Then the $Sp(4,\zed)$ matrix $\Lambda_b$ given by
\be\Lambda_b=\left(\begin{array}{cccc}U_1 & -\frac{n_2'}{r} & 0 & 0 \\ U_2 &
    \frac{n_1'}{r} & 0 & 0 \\ 0 & 0 & \frac{n_1'}{r} & -U_2 \\ 0 & 0 &
    \frac{n_2'}{r} & U_1 \end{array}\right)\ee
will perform the necessary operation. It belongs to an $SL(2,\zed)$
subgroup of the mapping class group. The desired transformation of
$\homo$ for which all temperature windings have been transfered to the
$a_1$ homology cycle is therefore described by the matrix
\beq
\homo':=\homo\,\Lambda_a\,\Lambda_b=\left(\begin{array}{cccc}r&0&0&0\\
x'&y'&z'&w \end{array}\right)
\label{homoprime}\eeq
where $x'=U_1\,r_1'+U_2\,r_2'$, $y'=\frac1r\,(n_1'\,r_2'-n_2'\,r_1')$,
$z'=\frac1r\,(n_1'\,s_1'+n_2'\,s_2')$ and $w=U_1\,s_2'-U_2\,s_1'$.

We now construct a third transformation matrix $\Lambda_c\in
Sp(4,\zed)$ by disregarding the first and third columns of the matrix
(\ref{homoprime}) and writing
\beq
\left(\begin{array}{cc} 0&0 \\ y'&w\end{array}\right)\,
\left(\begin{array}{cc} Y&-\frac wz \\ W&\frac{y'}z\end{array}\right)
=\left(\begin{array}{cc} 0&0 \\ z&0\end{array}\right) \ ,
\label{disregard}\eeq
where the integers $Y$ and $W$ obey
\beq
Y\,y'+W\,w={\rm gcd}(y',w)=:z \ .
\label{YWdef}\eeq
This does not affect the zeroes in the first row of (\ref{homoprime}),
and the symplectic matrix ${\sf T}=\Lambda_a\,\Lambda_b\,\Lambda_c$
finally reduces the matrix $\sf M$ to the form
\begin{equation}\label{partialreduction}
\homo'~\longmapsto~\left(\begin{array}{cccc} r & 0 & 0& 0 \\ x & y & z & 0
    \end{array}\right)
\end{equation}
with $x,y,z\in\zed$. Note that we do not apply the $2\times2$ matrix
$\sf S$ here, which affects an $SL(2,\zed)$ modular transformation of
the base $\torus^2_{\ii\nu}$. The complete Poincar\'e normal form
(\ref{PNF}) is derived in Appendix~A.

In this way the constraint equation
(\ref{matrixconstraint}) reduces to
\begin{equation}\pull^\top\,(\id_2,\Omega)=(1,\ii\nu)\,
\left(\begin{array}{cccc}
      r & 0 & 0& 0 \\ x & y & z & 0
    \end{array}\right)\,{\sf T} \ . \end{equation}
Now we factor out a symplectic unit on the right-hand side of this
equation in order that the eventual solution of the constraint
equation gives a period matrix with rational-valued off-diagonal
elements. This gives
\begin{equation}\label{readytoinvert}\pull^\top\,(\id_2,\Omega)=(1,\ii\nu)
\,\left(\begin{array}{cccc} 0&0&-r&0 \\z&0&-x&-y
    \end{array}\right)\,{\sf J}_2\,{\sf T} \ .\end{equation}
The matrix ${\sf J}_{2}\,{\sf T}\in Sp(4,\zed)$ is non-singular, and its
inverse $({\sf J}_{2}\,{\sf T})^{-1}$ acts on the left-hand side of
(\ref{readytoinvert}) as a modular transformation of the period matrix
$\Omega$ and the pullback vector $\pull^\top$. Parametrizing it by a
block matrix of the form (\ref{modulargroup}), one has
\beq
\pull^\top\,(\id_2,\Omega)\,({\sf J}_{2}\,{\sf T})^{-1}=
\pull^\top(C\,\Omega
+D)\,\Bigl(\id_2\,,\,(C\,\Omega +D)^{-1}\,(A\,\Omega +B)
\Bigr)=:\pull'\,\left(\id_2,\Omega '\,\right)
\label{Tabsorb}\eeq
giving
\be \label{finalconst}
\pull'\,\left(\id_2,\Omega'\,\right)=(1,\ii\nu)\,
\left(\begin{array}{cccc} 0&0&-r&0 \\z&0&-x&-y
    \end{array}\right) \ .\ee

We can now solve the constraint (\ref{finalconst}) to get
\beq
\pull=(1,\ii\nu)\,\left(\begin{array}{cc}0 & 0 \\ z
    & 0 \end{array}\right)=(\ii\nu\,z,0)
\eeq
and
\beq
\pull\,\Omega=(\ii\nu\,z,0)\,
\left(\begin{array}{cc}\Omega_{11} & \Omega_{12} \\ \Omega_{12} &
    \Omega_{22} \end{array}\right)=(1,\ii\nu)\,
\left(\begin{array}{cc}-r & 0 \\ -x &
    -y\end{array}\right) \ ,
\label{HOmegaeq}\eeq
where for notational ease we have dropped the primes indicating the
modular transformations (The free energy is modular invariant). The
period matrix is finally given in the form
\be\Omega\label{simpleperiod}=\left(\begin{array}{cc}-
\frac{x+\frac{r}{\ii\nu}}{z}& -\frac{y}{z} \\ -\frac{y}{z} &
    \Omega_{22} \end{array}\right) \ee
with $r,x,y,z\in\zed$ and $\Omega_{22}\in{\cal H}_1$. This form of the
period matrix has a natural geometrical interpretation. The diagonal
elements are related to the moduli of two tori which have been sewn
together along the branch cut of $\torus_{\ii\nu}^2$ to form the genus two
cover. The element $-\frac{x+r/\ii\nu}{z}$ is the modulus of a degree
$N=r\,z$ unbranched cover $\Sigma_1$ of the torus $\torus^2_{\ii\nu}$
as obtained in Section~\ref{OneLoop}. The off-diagonal element is a
measure of the radius and length of the cylinder joining the two tori
when they are glued together along the branch cut of $\torus_{\ii\nu}^2$. This
picture will be elucidated later on when we study degeneration limits
of the branched covers $\Sigma_2$ in Section~\ref{DegLimits}. Using
the projective modular symmetry $PSp(4,\zed)$ defining the moduli
space ${\cal M}_2$, we will identify $\Omega\sim-\Omega$ in
(\ref{simpleperiod}).

This calculation demonstrates that the existence of the
covering $f:\Sigma_2\to\torus^2_{\ii\nu}$, reducing a holomorphic
differential on $\Sigma_2$ to an elliptic one (\ref{pulldef}),
necessarily implies~\cite{BelEn1} the existence of another (generally
distinct) covering $f':\Sigma_2\to\torus^2_{\tau}$ which leads to a
reduction of some other independent holomorphic differential on
$\Sigma_2$ to an elliptic one. In this case, the Jacobian of the curve
$\Sigma_2$ represents a fibration whose base and fibre are the
elliptic curves $\torus_{\ii\nu}^2$ and $\torus_{\tau}^2$, with the
commutative diagram
\beq
\xymatrix{ & \torus^2_{\tau} \\
\Sigma_2 \ar[r]^{\!\!\!\!\!\!\!\!\!\mbf{\mathfrak{A}}}
  \ar[ru]^{f'} \ar[rd]_f&
{\rm Jac}(\Sigma_2) \ . \ar[d]^\Psi \ar[u]_{\Psi'}
\\ & \torus_{\ii\nu}^2 }
\label{extracommdiag}\eeq
The curve $\Sigma_2$ is embedded by the Abel map $\mathfrak{A}$ into
its Jacobian variety as a divisor. The relationship
(\ref{extracommdiag}) will then split the contribution to the two-loop
effective string action from $\Sigma_2$ into individual contributions
from the two tori $\torus^2_\tau$ and $\torus^2_{\ii\nu}$, as we shall
see explicitly below.

\subsection{Moduli Space}

The subgroup ${\cal G}$ of $Sp(4,\zed)$ transformations which
leave the structure of the integral matrices
\beq
\left(\begin{array}{cccc}0&0&-r&0\\
z&0&-x&-y\end{array}\right)
\eeq
in (\ref{finalconst}) invariant has four generators and consists of
unimodular matrices of the generic form
\begin{equation}\label{simplemodular}\left(\begin{array}{cccc}1 &
      A_{12} & B_{11} & B_{12} \\
        0& A_{22} & B_{12} & B_{22} \\
        0&0&1&0 \\
        0&C_{22}&D_{21}&D_{22}
      \end{array}\right)\end{equation}
which obey the non-linear constraints
\bea A_{22}\,D_{22}-B_{22}\,C_{22}&=&1 \ , \nonumber\\
A_{22}\,D_{21}-B_{21}\,C_{22}&=&A_{12} \ , \nonumber\\
B_{21}\,D_{22}-B_{22}\,D_{21}&=&B_{12} \ . \label{nonlinconstr}\eea
We choose $B_{11}$ and $B_{12}$ as arbitrary integers. From the Hopf
condition (\ref{Hopfcond}) it follows that the subgroup
${\cal G}$ preserves the two integers $r$ and $z$. Under a modular
transformation by an element (\ref{simplemodular}) of the group ${\cal
  G}$ the period matrix transforms according to
(\ref{Omegamodular}). By using (\ref{nonlinconstr}) one finds that the
matrix elements of $\Omega$ have the transformation properties
\bea
\Omega_{11}&\longmapsto&\Omega_{11}+B_{11}-\frac{C_{22}\,
(\Omega_{12})^2+2\,D_{21}\,\Omega_{12}+B_{12}\,D_{21}}
{C_{22}\,\Omega_{22}+{D_{22}}} \ , \nonumber\\
\Omega_{12}&\longmapsto&A_{22}\,\Omega_{12}+B_{21}-\frac{A_{22}\,
\Omega_{22}+B_{22}}{C_{22}\,\Omega_{22}+D_{22}}\,(C_{22}\,
\Omega_{12}+D_{21}) \ , \nonumber\\
\Omega_{22}&\longmapsto&\frac{A_{22}\,\Omega_{22}+B_{22}}{C_{22}\,
\Omega_{22}+D_{22}} \ .
\label{indtransflaws}\eea
Note that $\Omega_{22}$ transforms under a genus one $SL(2,\zed)$ modular
transformation. In addition the positivity of $\Omega_2$ yields the
quadratic constraints
\beq
{\rm Im}(\Omega_{11})>0 \ , \quad {\rm Im}(\Omega_{22})>0 \ , \quad
({\rm Im}\,\Omega_{12})^2<{\rm Im}(\Omega_{11})~{\rm Im}(\Omega_{22})
\ .
\label{Omega2pos}\eeq

{}From (\ref{indtransflaws}) and (\ref{Omega2pos}) it follows that the
period matrix take values in the extended fundamental
domain
\beq
{\cal F}_2'=\Bigl\{\Omega\in{\cal H}_2~\Bigm|~\Omega_{11}\in
\Delta'~,~\Omega_{22}\in\Delta~,~\Omega_{12}\in
{\cal P}_{\Omega_{22}}\Bigr\}
\label{calF2prime}\eeq
written in terms of the elliptic fundamental domains (\ref{calF1}) and
(\ref{calF1prime}) along with the parallelogram
\beq
{\cal
  P}_\tau=\Bigl\{\sigma_1+\tau\,\sigma_2~\Bigm|~\sigma_1,\sigma_2\in[0,1]
\Bigr\}
\label{parallelogram}\eeq
in the upper complex half-plane. The domain (\ref{calF2prime}) is the
same as the modular region obtained using the ordinary Rankin-Selberg
reduction~\cite{Moore1}. This provides a complete picture of
the moduli space of branched covers of a torus at
genus two. The map which sends a Riemann surface $\Sigma_2$ to
the equivalence class of the period matrix $\Omega\in{\cal M}_2$ is an
isomorphism onto the subspace ${\cal M}_2\setminus[{\cal
  H}_1\times{\cal H}_1]$, where
$[{\cal H}_1\times{\cal H}_1]$ is the modular orbit of the space of diagonal
period matrices in ${\cal H}_2$ corresponding to the boundary component of
moduli space where the surface $\Sigma_2$ degenerates into two
tori. The general task of finding an explicit set of inequalities on
the matrix elements of $\Omega$ which characterizes the corresponding
fundamental modular domain ${\cal F}_2$ is a difficult highly
non-linear mathematical problem. Here an explicit representation of
moduli space has been obtained by using reduction and unfolding
techniques. This is the main motivation behind our modification of the
Poincar\'e normal form, as it leads to a much simpler and tractable
fundamental modular region. For
completeness, the complete moduli space corresponding to the fully
reduced Poincar\'e normal form (\ref{PNF}) at genus two is worked out
explicitly in Appendix~A.

The sums over the integers in (\ref{calZg}) which characterize the
branched covers are restricted by the requirement that they count only
the moduli (\ref{simpleperiod}) lying in the extended fundamental domain
(\ref{calF2prime}). The positivity constraint (\ref{Omega2pos}) and
the Hopf condition (\ref{Hopfcond}) for the degree $N$ of the covering
map require $r,z\in\nat$ such that $r\,z=N$. The two equivalence relations
$\Omega_{12}\sim\Omega_{12}+1$ and $\Omega_{11}\sim\Omega_{11}+1$
imply that $x,y\in\zed/z\,\zed$, with $y\neq0$ in order for the period
matrix in ${\cal M}_2\setminus[{\cal H}_1\times{\cal H}_1]$ to
correspond to a genus two curve $\Sigma_2$. The arbitrary complex number
$\tau:=-\Omega_{22}\in{\cal H}_1$ is integrated over the genus
one fundamental domain $\Delta$. These ranges are all defined so that the
modular orbit of diagonal period matrices is removed from ${\cal
  H}_2$.

\subsection{Theta-Constants\label{ThetaFuns}}

The genus two theta-function with characteristics $\Theta:{\rm
  Jac}(\Sigma_2)\times{\cal H}_2\to\complex$ is defined as the Fourier
series~\cite{Mumford1}
\be\Theta{\mbf{a}\choose
    \mbf{b}}(\mbf{z}|\Omega)=\sum_{\mbf{n}\in\zed^2}\,
\exp\left[\pi\ii\left(\mbf{n}+\mbox{$\frac12$}
\,{\mbf{a}}\right)\cdot\Omega\left(\mbf{n}+\mbox{$\frac12$}\,
{\mbf{a}}\right)+2\pi\ii\left(\mbf{n}+\mbox{$\frac12$}\,
{\mbf{a}}\right)\cdot\left(\mbf{z}+\mbox{$\frac12$}\,{\mbf{b}}\right)
\right] \ .
\label{g2thetafn}\ee
It is a holomorphic function of
$(\mbf{z}|\Omega)\in\complex^2\times{\cal H}_2$ for the
characteristics $\mbf{a},\mbf{b}\in\real^2$. For $\mbf a,\mbf
b\in\{0,1\}\times\{0,1\}$ the theta-function is
even if $\mbf{a}\cdot\mbf{b}\equiv 0~\rm{mod}~2$, odd if
$\mbf{a}\cdot\mbf{b}\equiv 1~\rm{mod}~2$. There are ten even genus two
theta-functions and six odd ones. We can write (\ref{g2thetafn}) in a
form without characteristics by factorizing a phase to get
\be\label{Thetaphase}\Theta{\mbf{a}\choose
\mbf{b}}(\mbf{z}|\Omega)=\e^{\frac{\pi\ii}4\,{\mbf{a}\cdot
\Omega\mbf{a}}+\pi\ii\mbf{a}\cdot(\mbf{z}+\frac12\,{\mbf{b}})}~
\Theta{\mbf{0}\choose\mbf{0}}\left.\left(\mbf{z}+\mbox{$\frac12$}\,
{\Omega\mbf{a}}+\mbox{$\frac12$}\,{\mbf{b}}\right|\Omega\right)\ee

We can now use the reduction (\ref{simpleperiod}) to decompose the
genus two theta-function (\ref{g2thetafn}) in terms
of genus one theta-functions~\cite{Martens1}. The exponent of the
theta-function with zero characteristic in (\ref{Thetaphase}) is given
by the quantity
\bea
\mbf{k}\cdot\Omega\mbf k+2\,\mbf{k}\cdot\left(\mbf{z}+
\mbox{$\frac12$}\,\Omega\mbf{a}+\mbox{$\frac12$}\,\mbf{b}\right)
&=&(k_1)^2\,\Omega_{11}+2\,k_1\,\left(z_1+k_2\,\Omega_{12}+
\mbox{$\frac12$}\,\Omega_{11}\,
a_1+\mbox{$\frac12$}\,\Omega_{12}\,a_2+\mbox{$\frac12$}\,b_1\right)
\nonumber\\ &&+\,
(k_2)^2\,\Omega_{22}+2\,k_2\,\left(z_2+\mbox{$\frac12$}\,\Omega_{12}\,
a_1+\mbox{$\frac12$}\,\Omega_{22}\,a_2+\mbox{$\frac12$}\,b_2\right)
\label{fab}\eea
for $\mbf k=(k_1,k_2)\in\zed^2$. In the present case the period matrix
(\ref{simpleperiod}) (after projective $\zed_2$ reflection) has
rational-valued off-diagonal entries
$\Omega_{12}=\frac{y}{z}=\frac{r\,y}{N}$. Let $k_2=n+N\,\tilde{k}_2$
where $\tilde k_2\in\zed$ and $0\leq n \leq N-1$. We may then rewrite
(\ref{fab}) in the form
\bea
\mbf{k}\cdot\Omega\mbf k+2\,\mbf{k}\cdot\left(\mbf{z}+
\mbox{$\frac12$}\,\Omega\mbf{a}+\mbox{$\frac12$}\,\mbf{b}\right)
&=&(k_1)^2\,\Omega_{11}+2\,k_1\,\left[z_1+\left(n+
\mbox{$\frac{a_2}2$}\right)\,\Omega_{12}+\Omega_{11}\,
\mbox{$\frac{a_1}{2}$}+\mbox{$\frac{b_1}{2}$}\right]\nonumber\\
&&+\,2\,N\,\tilde{k}_2\,
\Omega_{12}+\left(\mbox{$\frac{n}{N}$}+\tilde{k}_2\right)^2\,
N^2\,\Omega_{22}\nonumber\\ && +\,2\,N\,\left(
\mbox{$\frac{n}{N}$}+\tilde{k}_2\right)\,\left(z_2+\Omega_{12}\,
\mbox{$\frac{a_1}{2}$}+\Omega_{22}\,\mbox{$\frac{a_2}{2}$}
+\mbox{$\frac{b_2}{2}$}\right) \ .
\label{fabfact}\eea
Once this expression is multiplied by $\pi\ii$ and exponentiated, the
term $2\pi \ii N\,\tilde{k}_2\,\Omega_{12}$ can be dropped since it is
an integer multiple of $2\pi\ii$. In this way the genus two theta-function
factorizes into elliptic theta-functions (\ref{thetagenus1}) as
\bea
\Theta{\mbf{a}\choose\mbf{b}}(\mbf{z}|\Omega)
  &=&\e^{-{\pi\ii a_1\,a_2\,r\,y}/{2\,N}}\,
  \sum_{n=0}^{N-1}\,\e^{-\pi\ii a_1\,n\,r\,y/N}~\theta{a_1 \choose
  b_1}\left(z_1+\big(n+\mbox{$\frac{a_2}{2}$}\big)\,\mbox{$
\frac{r\,y}N$}\left|\mbox{$\frac{r\,x+\frac{r^2}{\ii\nu}}N$}\right.\right)
\nonumber\\ && \qquad\qquad\qquad
\times\,\theta{\frac{2\,n+a_2}{N} \choose 0}\left.\left(N\,\big(z_2+\mbox{$
\frac{a_1\,r\,y}{2\,N}+\frac{b_2}{2}$}\big)\right|N^2\,\tau\right) \ .
\label{genus2thetafact}\eea
Each term in the sum over $n$ in (\ref{genus2thetafact}) contains a
pair of theta-functions, one for each of the tori in
(\ref{extracommdiag}).

Let us now restrict to theta-constants by setting $\mbf z=\mbf 0$ and
to integer characteristics $\mbf a,\mbf b\in\{0,1\}\times\{0,1\}$. The
decomposition (\ref{genus2thetafact}) into genus one theta-functions
can then be simplified somewhat by applying a Poisson
resummation to the second set of theta-functions with fractional
characteristics. For those characteristics, this results in the
modular transformation property~\cite{BatErd1}
\be
\theta{a\choose b}(z|\tau)=
\frac{\e^{-\pi\ii\big(\frac{z^2}{\tau}+\frac{a\,b}{2}\big)}}
{\sqrt{-\ii\tau}}~\theta{b\choose0}\left(\left.\mbox{$-\frac{a}{2}-
\frac{z}{\tau}$}\right|\mbox{$-\frac{1}{\tau}$}\right) \ .
\ee
Then the elliptic theta-functions are all given by standard integer
characteristic Jacobi-Erderlyi functions $\theta_a$ for $a=1,2,3,4$
and (\ref{genus2thetafact}) becomes
\beq
\Theta{\mbf{a}\choose\mbf{b}}(\mbf{0}|\Omega)=\frac{\e^{\pi\ii a_2\,
b_2/2}}{N\,\sqrt{-\ii\tau}}~\sum_{n=0}^{N-1}\,(-1)^{b_2\,n}~
\theta{a_1 \choose b_1}\left(\big(n+\mbox{$\frac{a_2}{2}$}\big)\,\mbox{$
\frac{r\,y}N$}\left|\mbox{$\frac{r\,x+\frac{r^2}{\ii\nu}}N$}\right.
\right)\,\theta_\gamma\left(\left.\mbox{$\frac{n+\frac{a_2}2}N$}\right|
-\mbox{$\frac1{N^2\,\tau}$}\right) \ ,
\label{JacErd2}\eeq
where $\gamma=2$ (resp. $\gamma=3$) when the degree $N$ and the
connecting integer $y$ are even (resp. odd).

\subsection{Free Energy\label{BosFreeen}}

We are finally ready to write down the genus two contribution to the
bosonic free energy. In the critical bosonic string theory, the
corresponding integration measure $\dd\mu_2^{\rm
  bos}(\Omega,\overline{\Omega}\,)$ on moduli space is completely
characterized by the fact that it is expressed in terms of the square
of a holomorphic volume form on ${\cal M}_2$ \cite{BelKn1,Morozov1,Moore1},
and by the requirement that it be free of global gravitational anomalies. It is
independent of temperature and has no zeroes or singularities in the interior
of moduli space, while on the boundary of moduli space (the $Sp(4,\zed)$ orbit
of ${\cal H}_1\times{\cal H}_1$ and infinity in ${\cal H}_2$) it has a second
order pole. This uniquely determines the bosonic integration measure as an
expansion in terms of modular forms. Using holomorphic factorization,
it can thereby be shown \cite{BelKn1,Morozov1,Moore1,DHPh1} that the
modular invariant, bosonic genus two moduli space measure is given by
\beq
\dd\mu_2^{\rm bos}\bigl(\Omega\,,\,\overline{\Omega}\,\bigr)
=\left(\frac{1}{4\pi^2\,\alpha'}\right)^{12}
{}~\dd^2\Omega_{11}\wedge\dd^2\Omega_{22}\wedge\dd^2
\Omega_{12}~\Bigl(\det \Omega_2
\Bigr)^{-13}~\Bigl|\Psi_{10}(\Omega)\Bigr|^{-2} \ ,
\label{dmu2bos}\eeq
where $\Psi_{10}$ is the unique, parabolic modular form of weight ten which
vanishes on the diagonal period matrices of ${\cal H}_2$ (This generalizes the
Dedekind function (\ref{Dedekind}) which comprises the one-loop moduli space
density for bosonic strings~\cite{Moore1}). It can be expressed in terms of the
ten even integer characteristic, genus two theta-constants as the
holomorphic Siegel cusp form
\be
\Psi_{10}(\Omega)=2^{-12}\,\prod_{\scriptstyle\mbf a\cdot\mbf b\,\equiv\,0\,
{{\rm mod}\,2}}~\left[\Theta{\mbf a\,\choose\mbf b\,}(\mbf0\,|\Omega)
\right]^2 \ .
\label{chi10}\ee

After integrating the delta-function in (\ref{calZg}) with the
necessary Jacobian from (\ref{HOmegaeq}) and (\ref{simpleperiod}), the
bosonic free energy is thereby found to be
\bea
F^{\rm bos}_2&=&-g^2_s\,\left(\frac1{2\,\sqrt2\,\pi\,\beta\,R}
\right)^{12}\,\sum_{N=1}^{\infty}
\,\frac{\e^{-\frac{\beta\,
      N}{\sqrt{2}\,R}}}{N^2}~\sum_{r\,z=N}\,
\left(\frac{z}{r}\right)^{10}\nn\\ && \times\,\sum_{\substack{
    x,y\in\zed/z\,\zed\\ y\neq0}}~
\int\limits_\Delta\,\frac{\dd^2\tau}{(\tau_2)^{12}}~
\prod_{\mbf a\cdot\mbf b\,\equiv\,0\,
{{\rm mod}\,2}}\,\Bigl|\Theta{\mbf a\,\choose\mbf b\,}
(\mbf{0}|\Omega)\Bigr|^{-4} \ .
\label{bosfreefinal}\eea
Using (\ref{JacErd2}) the product over genus two
theta-functions in (\ref{bosfreefinal}) can be expressed in terms of a
long string of integer characteristic Jacobi-Erderlyi functions
$\theta_a$, $a=1,2,3,4$. Generically these are {\it not}
theta-constants of the elliptic curves $\torus^2_{\ii\nu}$ and
$\torus_\tau^2$, as the connecting integers $y\in\zed/z\,\zed$ gluing
the two tori together appear in their arguments. The sums over the remaining
integers $N,r,z,x$ give the summation over worldsheet instanton sectors
$\Sigma_1\to\torus_{\ii\nu}^2$ characterizing the Hecke algebra. The
$\tau$-integral in (\ref{bosfreefinal}) gives the integration over the
location of the branch cut on the base $\torus^2_{\ii\nu}$ which is
used to construct the covering surface $\Sigma_2$ by gluing. This
identification can be established by using Thomae formulas to express
the branch points of the genus two curve transcendentally in terms of the
theta-constants (\ref{JacErd2})~\cite{BelEn1}. We will return to
these features in Section~\ref{DegLimits}.

As an aside, it is interesting to note that the cusp form
(\ref{chi10}) also arises in the computation~\cite{GritNik1} of the
elliptic genus of the Kummer surface ${\rm K}3$ as the one-loop free
energy of a single string given by
\beq
\chi_{{\rm K}3}^{~}(\zeta|\tau)=8\,\sum_{a=2}^4\,\frac{
\theta_a(\zeta|\tau)^2}{\theta_a(0|\tau)^2} \ ,
\label{chiK3}\eeq
where $\Omega_{12}=\zeta$ is the complexified K\"ahler form of the
elliptic curve $\torus^2_\tau$. The completion of the corresponding
string partition function on the symmetric product orbifold of ${\rm
  K}3$ to an automorphic form for the group $SO(3,2,\zed)$ is simply
(\ref{chi10}). This form can also be interpreted as the denominator of
a generalized Kac-Moody algebra~\cite{GritNik1,Kawai1}. Our reduction
formulas here and in what follows bear a remarkable similarity to this
construction, with the ${\rm K}3$ surface regarded as the resolution
of the orbifold $(\torus^2_{\ii\nu}\times\torus^2_\tau)/\zed_2$. It
would be interesting to further pursue whether or not our two-loop
partition functions admit deeper interpretations along these lines.

\newsection{Superstrings}

We now turn to our main object of interest, the two-loop superstring
free energy. There are two new ingredients in this case that one must
add to the calculation of the previous section. In the
genus one case, the simplicity of the measure (\ref{dmu1}) is a result of the
local cancellation between the longitudinal $X$ and ghost $B,C$ determinants on
moduli space. This ceases to happen for genera $g>1$, and in this case the
calculations are notoriously subtle. We shall take the standard prescription
for obtaining the measure $\dd\mu_2[{{\mbf n}\atop{\mbf
m}}](\Omega,\overline{\Omega}\,)$ by integrating over the
fermionic moduli~\cite{VV1}. The non-splitness of super-moduli space
does not generically allow for a
global, unambiguous reduction to ordinary moduli, since the Grassmann
integrations lead to spurious gauge dependences in the form of total derivative
terms on ${\cal M}_2$ \cite{AtMS1}. The problem can be overcome by
descending from super-moduli space to moduli space by projecting
super-geometries onto super-period matrices~\cite{DHPh1}. The integration over
Grassmann odd supermoduli is then performed by integrating over the
fibers of this projection. With this, one can find a good global
holomorphic gauge slice in Teichm\"uller space without spurious gauge
dependences that could otherwise lead to modular anomalies in the measure on
moduli space. For each even spin structure at $g=2$, slice-independence allows
an arbitrary choice of worldsheet gravitino field insertion
point~\cite{FalReina1} and the split gauge choice leads to an
expression for the chiral superstring measure in terms of modular
forms~\cite{DHPh1}. The contributions from
odd spin structures again vanish as a result of the integration over
fermionic zero modes. For fixed spin structure, the chiral measure
allows for a unique modular covariant GSO projection~\cite{DHPh1},
which must be appropriately modified~\cite{AtWitten1} due to the
finite-temperature supersymmetry breaking effects analogously to the
one-loop case.

\subsection{Spin Structures}

Let us begin by setting some useful shorthand notations. A reduced
genus two integer characteristic is a pair of vectors
$\bigl({\mbf{a}\atop\mbf{b}}\bigr)=\bigl({}^{a_1~b_1}_{a_2~b_2}\bigr)$ where
each
  $a_i,b_i$, $i=1,2$ are either $0$ or $1$.  A characteristic is even
  if $\mbf{a}\cdot\mbf{b}\equiv 0~\rm{mod}~2$, odd if
  $\mbf{a}\cdot\mbf{b}\equiv 1~\rm{mod}~2$. There are ten even
  characteristics and six odd characteristics associated to the
  distinct choices of spin structures on the genus two Riemann surface
  $\Sigma_2$, i.e. to the choices of a square root of the canonical
  line bundle over $\Sigma_2$. The ten even characteristics (spin
  structures) are denoted
\bea
&&\delta_1=\left({}^{00}_{00}\right)\qquad
\delta_2=\left({}^{00}_{01}\right)\qquad
\delta_3=\left({}^{01}_{00}\right)\qquad
\delta_4=\left({}^{01}_{01}\right)\qquad
\delta_5=\left({}^{00}_{10}\right) \nonumber\\
&&\delta_6=\left({}^{01}_{10}\right)\qquad
\delta_7=\left({}^{10}_{00}\right)\qquad
\delta_8=\left({}^{10}_{01}\right)\qquad
\delta_9=\left({}^{10}_{10}\right)\qquad
\delta_{0}=\left({}^{11}_{11}\right) \ .
\label{10evenchars}\eea
The odd spin structures are denoted
\bea
\nu_1=\left({}^{00}_{11}\right)\qquad
\nu_2=\left({}^{01}_{11}\right)\qquad
\nu_3=\left({}^{11}_{00}\right)\qquad
\nu_4=\left({}^{11}_{01}\right)\qquad
\nu_5=\left({}^{10}_{11}\right)\qquad
\nu_6=\left({}^{11}_{10}\right) \ .
\label{6oddchars}\eea
Integer characteristics may be summed modulo 2, componentwise as if they
were $2\times2$ matrices. For example, $\nu_1+\nu_4+\nu_6=\delta_0$
and $\nu_2+\nu_3+\nu_5=\delta_0$. There is a two-to-one map
from triples of odd characteristics which are pairwise distinct onto
even characteristics. The relative signature between any two spin
structures is defined by
\begin{equation}
\bigl\langle\bigl({}^{\mbf{a}}_{\mbf{b}}\bigr)
\big|\bigl({}^{\mbf{a}'}_{\mbf{b}'}\bigr)\bigr\rangle=
\exp\Bigl[\pi\ii\bigl(\mbf{a}\cdot\mbf{b}'-\mbf{b}\cdot\mbf{a}'\,
\bigr)\Bigr] \ .
\end{equation}

\subsection{GSO Projection\label{GSO}}

In order for spacetime fermions and spacetime bosons to have the
correct statistics at finite temperature, the fermions must have
antiperiodic boundary conditions and the bosons must be periodic
around the temperature direction of the target space. The standard GSO
projection is thus modified by phases which take into
account the winding numbers $\mbf n$ and $\mbf
m$~\cite{AtWitten1}. A genus two Riemann surface has 16 spin
structures given by the generators (\ref{10evenchars}) and
(\ref{6oddchars}) of the cohomology group
$H^1(\Sigma_2,\zed/2\,\zed)=(\zed/2\,\zed)^4$ which are in one-to-one
correspondence with flat real line bundles $L\to\Sigma_2$. Define
$\phi(L)=+1$ if the spin structure $L$ is even and $\phi(L)=-1$ if it
is odd. This quantity coincides with the mod~2 index~\cite{Mumford1}
\beq
\phi(L)=\exp\Bigl[\pi\ii\dim H^0\big(\Sigma_2\,,\,Spin(\Sigma_2)
\otimes L\big)\Bigr]
\label{phiLdef}\eeq
which counts the number of holomorphic sections of the twisted spinor
bundle $Spin(\Sigma_2)\otimes L$ modulo~2. The reduction modulo 2 of $(\mbf
n,\mbf m)$ defines the characteristic class in
$H^1(\Sigma_2,\zed/2\,\zed)$ of a flat connection of a real line
bundle ${\cal L}_{(\mbf n,\mbf m)}\to\Sigma_2$ such that a holomorphic
section of ${\cal L}_{(\mbf n,\mbf m)}$
changes by a phase $(-1)^{n_i}$ as one goes once around the $a_i$
homology cycles and by $(-1)^{m_i}$ as one goes once around the $b_i$
homology cycles. Given a spin structure $L$, the tensor product
$L\otimes{\cal L}_{(\mbf n,\mbf m)}$ is another spin structure for
any $\mbf n,\mbf m$ and we define
\be U_L(\mbf n,\mbf m)=\phi(L)\,\phi(L\otimes{\cal L}_{(\mbf n,\mbf
  m)}) \ .
\label{ULphase}\ee
The quantity (\ref{ULphase}) takes values $\pm\,1$, and it is the correct
phase to insert into the sum over spin structures $L$ and winding
numbers $\mbf n,\mbf m$~\cite{AtWitten1}. It is compatible with both
factorization of $\Sigma_2$ to lower genus and modular invariance.

As an example, let us calculate
$U_{\delta_5}(\mbf n,\mbf m)=U_{({}^{00}_{10})}(\mbf n,\mbf m)$. The
spin structure $\delta_5$ is even so
$U_{\delta_5}(\mbf n,\mbf m)=\phi(L_{\delta_5}\otimes{\cal L}_{(\mbf
  n,\mbf m)})$. We first calculate $U_{\delta_5}(n_1,0,0,0)$. For
$n_1\in\zed$ odd one has $L_{\delta_5}\otimes{\cal
  L}_{(n_1,0,0,0)}=L_{\delta_9}$ and so
$U_{\delta_5}(n_1,0,0,0)=\phi(L_{\delta_9})=1$. The spin structure
$L_{\delta_5}\otimes{\cal L}$ is likewise even if
$\cal L$ corresponds to wrapping the $a_2$ and $b_1$ cycles around the
temperature direction odd numbers of times $n_2$ and $m_1$, and so we
have
\beq
U_{\delta_5}(n_1,0,0,0)=U_{\delta_5}(0,0,m_1,0)=U_{\delta_5}
(0,n_2,0,0)=1 \ .
\label{1non0}\eeq
When the $b_2$ cycle wraps around the temperature direction an odd
number of times $m_2$ one gets the spin structure
$L_{\delta_5}\otimes{\cal L}_{(0,0,0,m_2)}=L_{\nu_1}$. The
phase is then
\beq
U_{\delta_5}(0,0,0,m_2)=(-1)^{m_2} \ .
\label{m2non0}\eeq

Taking into account pairs of cycles produces the phase factors
\bea
U_{\delta_5}(0,n_2,m_1,0)&=&1 \ ,
\nonumber\\
U_{\delta_5}(0,0,m_1,m_2)&=&U_{\delta_5}(n_1,0,0,m_2)~=~(-1)^{m_2} \ ,
\nonumber\\
U_{\delta_5}(n_1,0,m_1,0)&=&\mbox{$\frac{1}{2}$}\,
\left(1+(-1)^{n_1}+(-1)^{m_1}-(-1)^{n_1+m_1}\right) \ ,
\nonumber\\
U_{\delta_5}(0,n_2,0,m_2)&=&\mbox{$\frac{1}{2}$}\,
\left(1-(-1)^{n_2}+(-1)^{m_2}+(-1)^{n_2+m_2}\right) \ .
\label{2non0}\eea
For triples of cycles the phases are given by
\bea
U_{\delta_5}(n_1,0,m_1,m_2)&=&\mbox{$\frac{1}{2}$}\,
\left((-1)^{m_2}+(-1)^{n_1+m_2}+(-1)^{m_1+m_2}-(-1)^{n_1+m_1+m_2}
\right) \ , \nonumber\\
U_{\delta_5}(0,n_2,m_1,m_2)&=&U_{\delta_5}(0,n_2,0,m_2) \ , \nonumber\\
U_{\delta_5}(n_1,n_2,m_1,0)&=&U_{\delta_5}(n_1,0,m_1,0) \ ,
\nonumber\\
U_{\delta_5}(n_1,n_2,0,m_2)&=&U_{\delta_5}(0,0,n_2,m_2) \ .
\label{3non0}\eea
The GSO phases for any $n_i,m_i$, $i=1,2$ are given generally by an
expression of the form
\bea
U_{\delta_5}(\mbf n,\mbf m)&=&\alpha\,(-1)^{n_1+n_2+m_1+m_2}+
\beta_1\,(-1)^{n_1+n_2+m_1}+
\beta_2\,(-1)^{n_1+n_2+m_2}+\beta_3\,(-1)^{n_1+m_1+m_2}\nonumber\\
  &&+\,\beta_4\,(-1)^{n_2+m_1+m_2}+\gamma_1\,(-1)^{n_1+n_2}+\gamma_2\,
(-1)^{m_1+m_2}+\gamma_3\,(-1)^{n_1+m_1}\nonumber\\
  &&+\,\gamma_4\,(-1)^{n_2+m_2}+\gamma_5\,(-1)^{n_1+m_2}+
\gamma_6\,(-1)^{n_2+m_1}+\varepsilon_1\,
(-1)^{n_1}+\varepsilon_2\,(-1)^{n_2}\nonumber\\
&&+\,\varepsilon_3\,(-1)^{m_1}+\varepsilon_4\,(-1)^{m_2}+\eta \ .
\label{allnon0}\eea
The phase (\ref{allnon0}) must reduce to (\ref{1non0})--(\ref{3non0})
when the appropriate winding numbers are set to zero. This gives a
system of equations which are enough to determine $U_{\delta_5}(\mbf
n,\mbf m)$ up to a proportionality constant which may be fixed by
requiring the phase to be normalised as $\pm\,1$.

One can compute all 16 phase factors in this way as functions of
generic thermal winding numbers $\mbf n,\mbf m$. After some inspection
and calculation, one finds that as a function
$U:\{0,1\}^2\times\zed\to\{\pm\,1\}$ the GSO phase (\ref{ULphase})
is given by
\bea
U_{({}^{\mbf a}_{\mbf b})}(\mbf n,\mbf m)&=&\mbox{$\frac14$}\,
(-1)^{\mbf a\cdot\mbf b}\,
\Bigl[(-1)^{n_1+n_2+m_1+m_2}\,(-1)^{a_1+a_2+b_1+b_2}-
(-1)^{n_1+n_2+m_1}\,(-1)^{a_1+a_2+b_1}\Bigr.\nonumber\\ &&
-\,(-1)^{n_1+n_2+m_2}\,(-1)^{a_1+a_2+b_2}-(-1)^{n_1+m_1+m_2}\,
(-1)^{a_1+b_1+b_2}\nonumber\\ && -\,(-1)^{n_2+m_1+m_2}\,(-1)^{a_2+b_1+b_2}
+(-1)^{n_1+n_2}\,(-1)^{a_1+a_2}+(-1)^{m_1+m_2}\,
(-1)^{b_1+b_2}\nonumber\\ &&-\,(-1)^{n_1+m_1}\,(-1)^{a_1+b_1}
-(-1)^{n_2+m_2}\,(-1)^{a_2+b_2}+(-1)^{n_1+m_2}\,(-1)^{a_1+b_2}
\nonumber\\ &&+\,(-1)^{n_2+m_1}\,(-1)^{a_2+b_1} -
(-1)^{n_1}\,(-1)^{a_1}+(-1)^{n_2}\,(-1)^{a_2}\nonumber\\ &&
+\Bigl.\,(-1)^{m_1}\,(-1)^{b_1}
+(-1)^{m_2}\,(-1)^{b_2}+1\Bigr] \ .
\label{GSOphasegen}\eea
For our particular calculation the Riemann surface $\Sigma_2$ is a branched
covering of a torus, and the modified GSO projection is very simple since
there is only one homology cycle of the cover which is wrapped around the
temperature direction after the reduction to (\ref{partialreduction})
given by $(\mbf n,\mbf m)\to (0,0,r,0)$. The only even spin structure
GSO phases which are non-trivial for generic $r$ are given by
\beq
U_{\delta_7}(0,0,r,0)=U_{\delta_8}(0,0,r,0)=
U_{\delta_9}(0,0,r,0)=U_{\delta_0}(0,0,r,0)=(-1)^r \ .
\label{Udeltanontriv}\eeq
All other even spin structure phases are equal to~$1$.

\subsection{Chiral Measure}

Holomorphic factorization of the genus two superstring measure at zero
temperature is achieved by trading the Belavin-Knizhnik obstruction
(encoded through the partition function of a
free chiral scalar field on $\Sigma_2$ given by
$(4\pi^2\,\alpha'\,)^{-5}\,(\det\overline{\partial}{}_0)^{-10}$) for
an integral over internal loop momenta $\mbf p_\mu\in\real^2$,
$\mu=0,1,\dots,9$ flowing through the $\mbf a$ cycles of
$\Sigma_2$~\cite{VV1}. The resulting dependence on moduli and spin
structures is intricately encoded into various sections of the twisted
spinor bundles over $\Sigma_2$~\cite{VV2,EOo1}, which may be expressed
in terms of modular forms associated with the Riemann
surface~\cite{DHPh1}. The {\it chiral} measure corresponding to a
fixed even spin structure $\delta$ is required to be free of global
gravitational anomalies on super-moduli space before integrating out
the fermionic moduli~\cite{DHPhrev}. On ${\cal M}_2$ it may be
computed explicitly to be~\cite{DHPh1}
\begin{equation}
\dd\mu_2[\delta](\Omega)=\left(\frac{1}{4\pi^2\,\alpha'}
\right)^2~\dd\Omega_{11}\wedge
\dd\Omega_{22}\wedge\dd\Omega_{12}~\frac{\Xi_6[\delta](\Omega)\,
\Theta[\delta](\mbf 0|\Omega)^4}{\Psi_{10}(\Omega)}  \ ,
\label{chiralmeas0temp}\end{equation}
where $\Psi_{10}$ is the modular form of weight ten defined in
(\ref{chi10}) which arises as the bosonic contribution, and
\begin{equation}
  \Xi_6[\delta](\Omega):=
\sum_{1\leq k< l\leq 3}\,\langle\nu_{i_k}|\nu_{i_l}\rangle\,
\prod_{j=4,5,6}\Theta[\nu_{i_k}+\nu_{i_l}+\nu_{i_j}](\mbf 0|\Omega)^4 \ .
\label{Xi6def}\end{equation}
Here we have chosen a partition
$\{i_1,i_2,i_3\}\cup\{i_4,i_5,i_6\}=\{1,2,3,4,5,6\}$ of the index set
labelling the odd characteristics (\ref{6oddchars}) such that
$\delta=\nu_{i_1}+\nu_{i_2}+\nu_{i_3}=\nu_{i_4}+\nu_{i_5}+\nu_{i_6}$. The
quantity
(\ref{Xi6def}) depends only on the spin structure $\delta$ and not on
the actual triplet of odd characteristics used to represent
$\delta$. This follows from the fact that the odd spin structures
$\nu_{i_1},\nu_{i_2},\nu_{i_3}$ result from a choice of worldsheet
gravitino field insertion points, and the two-loop chiral measure is
completely independent of these points~\cite{DHPh1}. The object
$\Xi_6[\delta](\Omega)$ has modular weight~$6$, but it is {\it not} a
modular form because it depends on the spin structure $\delta$ and an
additional sign factor arises in its modular transformation
laws~\cite{DHPh1}. As a consequence, the measure
(\ref{chiralmeas0temp}) is modular covariant of weight~$-5$.

The full chiral genus two superstring measure is obtained
by summing (\ref{chiralmeas0temp}) over all even spin structures
$\delta$ with weights provided by the phases $U_\delta(\mbf n,\mbf m)$
computed in Section~\ref{GSO} which take into account the modification
of the GSO projection at finite temperature. Thus we define
\begin{equation}\label{finitesupermeasure}
  \dd\mu_2\left[{\mbf n}\atop{\mbf m}\right]
(\Omega)=\sum_{\delta\,{\rm even}}\, U_\delta(\mbf n,\mbf m)~
\dd\mu_2[\delta](\Omega) \ .
\end{equation}
The quantity $\Upsilon_8(\Omega):=\sum_{\delta\,{\rm
even}}\,\Xi_6[\delta](\Omega)\,
\Theta[\delta](\mbf 0|\Omega)^4$ is a uniquely constructed modular
form of weight~$8$. Using the Riemann bilinear relations one can show
that~\cite{DHPh1}
$\Upsilon_8(\Omega)=2\,\Psi_8(\Omega)-\frac12\,\Psi_4(\Omega)^2$ where
$\Psi_8(\Omega)$ and $\Psi_4(\Omega)$ are respectively the weight~$8$
and weight~$4$ generators of the polynomial ring of genus two modular
forms. By Igusa's theorem~\cite{Igusa2}, $\Psi_8(\Omega)$ is the
unique modular form of weight~$8$ with
$4\,\Psi_8(\Omega)=\Psi_4(\Omega)^2$. It follows that
$\Upsilon_8(\Omega)=0$ and thus we have the identity
\beq
\sum_{\delta\,{\rm even}}\,\Xi_6[\delta](\Omega)\,
\Theta[\delta](\mbf 0|\Omega)^4=0 \ .
\label{Upsilon8id}\eeq
Using this along with the modified GSO phases (\ref{Udeltanontriv})
corresponding to the reduced form of the period matrix
(\ref{simpleperiod}), we can bring (\ref{finitesupermeasure}) to the
form
\bea
\dd\mu_2\left[{{}^0_0}\atop{{}^r_0}
\right](\Omega)&=&\left(\frac{1}{4\pi^2\,\alpha'}\right)^2~
\dd\Omega_{11}\wedge\dd\Omega_{22}
\wedge\dd\Omega_{12}~\frac{\bigl(\e^{\pi\ii r}-1\bigr)}{\Psi_{10}(\Omega)}
\nonumber\\ &&\times\,\Bigl(\Xi_6[\delta_7](\Omega)\,\Theta[\delta_7]
(\mbf 0|\Omega)^4+\Xi_6[\delta_8](\Omega)\,\Theta[\delta_8](\mbf 0|
\Omega)^4\Bigr.\nonumber\\ &&\qquad+\Bigl.\,
\Xi_6[\delta_9](\Omega)\,\Theta[\delta_9](\mbf 0|\Omega)^4+
\Xi_6[\delta_0](\Omega)\,\Theta[\delta_0](\mbf 0|\Omega)^4\Bigr) \ .
\label{modmeasure}\eea
As in the one-loop case, when $r$ is even the Fermi fields are periodic
and so the fermions and bosons have the same boundary conditions. These
sectors are supersymmetric, and the mode expansions of both the fermion
and boson fields contain zero modes. The integration over fermionic
zero modes gives zero. Hence (\ref{modmeasure}) vanishes, as expected
in the supersymmetric sectors.

\subsection{Free Energy\label{SUSYFreeen}}

The chiral measure (\ref{finitesupermeasure}) is a modular form of
weight $-5$. When we include both left and right moving degrees of
freedom of the string theory, the non-chiral measure $\dd\mu[{{\mbf
  n}\atop{\mbf m}}](\Omega)\wedge\overline{\dd\mu[{{\mbf n}\atop{\mbf
    m}}](\Omega)}$ is a modular form of weight $-10$. The complete
measure which defines a modular invariant function on moduli space
${\cal M}_2$ is thus
\beq
\dd\mu_2\left[{\mbf n}\atop{\mbf m}\right]\bigl(\Omega\,,\,
\overline{\Omega}\,\bigr)=\bigl(\det\Omega_2)^{-5}~\dd\mu_2\left[{\mbf
    n}\atop{\mbf m}\right](\Omega)\wedge
\overline{\dd\mu_2\left[{\mbf n}\atop{\mbf m}\right](\Omega)} \ .
\label{modinvsusymeas}\eeq

We now substitute the full superstring measure (\ref{modinvsusymeas})
into (\ref{calZg}) using (\ref{modmeasure}), and resolve the
delta-function constraint after performing the necessary reduction to
(\ref{simpleperiod}) (including the appropriate Jacobian). The
superstring free energy is thereby found to
be
\bea
F_2&=&-\frac{g_s^2}{4}\,\left(\frac1{4\,\sqrt2\,\pi\,\beta\,R}
\right)^4\,\sum_{N=1}^{\infty}\,
\frac{\e^{-\frac{\beta\,
      N}{\sqrt{2}\,R}}}{N}~\sum_{\substack{r\,z=N\\ r
    \,\textrm{odd}}}\,\frac{1}{r^4}~
\sum_{\substack{x,y\in\zed/z\,\zed\\
y\neq0}}~\int\limits_{\Delta}
\frac{\dd^2\tau}{(\tau_2)^4}~\Bigl|\Psi_{10}(\Omega)\Bigr|^{-2}
\nonumber\\ && \times\,\Bigl|\Xi_6[\delta_7](\Omega)\,\Theta[\delta_7]
(\mbf 0|\Omega)^4+\Xi_6[\delta_8](\Omega)\,\Theta[\delta_8](\mbf 0|
\Omega)^4\Bigr.\nonumber\\ &&\qquad+\Bigl.\,
\Xi_6[\delta_9](\Omega)\,\Theta[\delta_9](\mbf 0|\Omega)^4+
\Xi_6[\delta_0](\Omega)\,\Theta[\delta_0](\mbf 0|\Omega)^4
\Bigr|^2 \ .
\label{finalfreeenergy}\eea
The quantities (\ref{Xi6def}) are worked out in Appendix~B. We denote
$\Theta_i(\Omega):=\Theta[\delta_i](\mbf 0|\Omega)$. Using the
explicit expression (\ref{chi10}), the integrand in
(\ref{finalfreeenergy}) can be expanded out in terms of the ten even
characteristic genus two theta-constants to get
\bea
F_2&=&-\frac{g_s^2}{4}\,\left(\frac1{4\,\sqrt2\,\pi\,\beta\,R}
\right)^4\,\sum_{N=1}^{\infty}\,
\frac{\e^{-\frac{\beta\,
      N}{\sqrt{2}\,R}}}{N}~\sum_{\substack{r\,z=N\\ r
    \,\textrm{odd}}}\,\frac1{r^4}\nn\\ && \times\,
\sum_{\substack{x,y\in\zed/z\,\zed\\
y\neq0}}~\int\limits_{\Delta}
\frac{\dd^2\tau}{(\tau_2)^4}~\Bigg\arrowvert
4\,\left(\frac{\Theta_7\,\Theta_8\,\Theta_9\,\Theta_0}
{\Theta_1\,\Theta_2\,\Theta_3\,\Theta_4\,\Theta_5\,\Theta_6}\right)^2
 \Bigg. \nonumber\\ &&
  +\,\left(\frac{\Theta_2\,\Theta_3\,\Theta_5\,\Theta_7}
{\Theta_1\,\Theta_4\,\Theta_6\,\Theta_8\,\Theta_9\,\Theta_0}\right)^2
    -\left(\frac{\Theta_1\,\Theta_4\,\Theta_6\,\Theta_7}
{\Theta_2\,\Theta_3\,\Theta_5\,\Theta_8\,\Theta_9\,\Theta_0}\right)^2
+\left(\frac{\Theta_2\,\Theta_3\,\Theta_6\,\Theta_8}
{\Theta_1\,\Theta_4\,\Theta_5\,\Theta_7\,\Theta_9\,\Theta_0}\right)^2
  \nonumber\\ &&
      -\,\left(\frac{\Theta_1\,\Theta_4\,\Theta_5\,\Theta_8}
{\Theta_2\,\Theta_3\,\Theta_6\,\Theta_7\,\Theta_9\,\Theta_0}\right)^2
    +\left(\frac{\Theta_3\,\Theta_4\,\Theta_5\,\Theta_9}
{\Theta_1\,\Theta_2\,\Theta_6\,\Theta_7\,\Theta_8\,\Theta_0}\right)^2
-\left(\frac{\Theta_1\,\Theta_2\,\Theta_6\,\Theta_9}
{\Theta_3\,\Theta_4\,\Theta_5\,\Theta_7\,\Theta_8\,\Theta_0}\right)^2
  \nonumber\\ &&
      +\Bigg.\,\left(\frac{\Theta_3\,\Theta_4\,\Theta_6\,\Theta_0}
{\Theta_1\,\Theta_2\,\Theta_5\,\Theta_7\,\Theta_8\,\Theta_9}\right)^2
    -\left(\frac{\Theta_1\,\Theta_2\,\Theta_5\,\Theta_0}
{\Theta_3\,\Theta_4\,\Theta_6\,\Theta_7\,\Theta_8\,\Theta_9}\right)^2
\Bigg\arrowvert^2 \ .
\label{F2exptheta}\eea
The theta-constants appearing in (\ref{F2exptheta}) are functions of
the period matrix (\ref{simpleperiod}) and therefore depend on both
the discrete and continuous parameters which characterize the branched
covers~$\Sigma_2$. Their explicit forms in terms of elliptic
Jacobi-Erderlyi functions are given by the
formula~(\ref{JacErd2}). We have not found any genus one
theta-function identities which could simplify (\ref{F2exptheta}) and
make this expression more explicit.

Note that, in contrast to the one-loop case
which relied solely on the Jacobi identity (\ref{abstruse}), the
modular invariance of the two-loop thermodynamic free energy does not
follow from Riemann identities alone but in addition requires a
special property of the ring of modular forms at genus two. The
drastic difference between the summation prefactors in the bosonic
case (\ref{bosfreefinal}) and in the supersymmetric case
(\ref{F2exptheta}) reflects the different analytic natures
of the associated twist field perturbations described in
Section~\ref{Intro}. This difference will be encountered again in a
more explicit form in Section~\ref{Factorization}.

\newsection{Heterotic Strings\label{Heterotic}}

Let us now describe how our analysis applies to heterotic string
theory. We replace the matter field action
$S[X]+\overline{S[X]}$ in (\ref{Polyakovaction}) by
\beq
S_{\rm het}[X,\lambda]=\frac1{4\pi\,\alpha'}\,\int\limits_{\Sigma_2}\,
\dd^2z~\Bigl(|\partial x^\mu|^2+\psi_\mu\,\overline{\partial}\psi^\mu+
\lambda_A\,\partial\lambda^A\Bigr) \ ,
\label{heteroticaction}\eeq
where the fermionic fields $\lambda^A$, $A=1,\dots,32$ are Lorentz
singlets. Both $\psi^\mu$ and $\lambda^A$ are Majorana-Weyl fermion
fields. The ghost contributions are unchanged. Thus the left-moving
(holomorphic) part of the heterotic string coincides with that of the
superstring whose chiral modular covariant measure is given by
(\ref{modmeasure}). After bosonization of $\lambda^A$, the
right-moving (antiholomorphic) part coincides with that of the bosonic
string of Section~\ref{BosFreeen} with $16$ anti-chiral bosons
compactified on the Cartan torus of the heterotic gauge group $G$,
where $G={Spin(32)}/{\zed_2}$ or $G=E_8\times E_8$. The compactified
bosonic fields produce an extra winding contribution given by a
theta-function of the root lattice of $G$, which at genus two is the
unique modular form of weight eight given by~\cite{Moore1}
\beq
\Psi_8(\Omega)=\sum_{\delta\,{\rm even}}\,\Theta[\delta](\mbf 0|
\Omega)^{16} \ .
\label{Psi8def}\eeq
It follows that the two-loop anti-chiral heterotic string measure
is~\cite{DHPh1}
\beq
\dd\mu_2^{\rm het}\bigl(\,\overline{\Omega}\,\bigr)=
\left(\frac{1}{4\pi^{2}\,\alpha'}\right)^6
{}~\dd\overline{\Omega}{}_{11}\wedge\dd
\overline{\Omega}{}_{12}\wedge\dd\overline{\Omega}{}_{22}~
\frac{\overline{\Psi_8(\Omega)}}{\overline{\Psi_{10}(\Omega)}} \ .
\label{chiralhetmeas}\eeq

The full modular invariant non-chiral measure is thus
$(\det\Omega_2)^{-5}~\dd\mu_2[{{\mbf n}\atop{\mbf
  m}}](\Omega)\wedge\dd\mu_2^{\rm
  het}(\,\overline{\Omega}\,)$. Substituting this into (\ref{calZg})
using (\ref{modmeasure}) and (\ref{chiralhetmeas}), by proceeding as
before we find that the heterotic string free energy is given by
\bea
F^{\rm het}_{2}&=&\frac{g_s^2}{8}\,\left(\frac1{128\,\sqrt2\,
\pi^3\,\alpha'\,\beta\,R}\right)^4\,
\sum_{N=1}^{\infty}\,\frac{\e^{-\frac{\beta\,
      N}{\sqrt{2}\,R}}}{N}~\sum_{\substack{r\,z=N\\ r
    \,\textrm{odd}}}\,\frac1{r^4}~\sum_{\substack{x,y\in\zed/z\,\zed\\
y\neq0}}~\int\limits_{\Delta}
\frac{\dd^2\tau}{(\tau_2)^4}~\frac{\overline{\Psi_8(\Omega)}}{
\Bigl|\Psi_{10}(\Omega)\Bigr|^{2}}
\nonumber\\ && \times\,\Bigl(\Xi_6[\delta_7](\Omega)\,\Theta[\delta_7]
(\mbf 0|\Omega)^4+\Xi_6[\delta_8](\Omega)\,\Theta[\delta_8](\mbf 0|
\Omega)^4\Bigr.\nonumber\\ &&\qquad+\Bigl.\,
\Xi_6[\delta_9](\Omega)\,\Theta[\delta_9](\mbf 0|\Omega)^4+
\Xi_6[\delta_0](\Omega)\,\Theta[\delta_0](\mbf 0|\Omega)^4
\Bigr) \ .
\label{Heteroticfreeenergy}\eea
As in (\ref{F2exptheta}), this expression can be expanded into the ten
even characteristic genus two theta-constants by using (\ref{chi10}),
(\ref{Psi8def}) and the formulas of Appendix~B to get
\bea
F_2^{\rm het}&=&\frac{g_s^2}{8}\,\left(\frac1{128\,\sqrt2\,
\pi^3\,\alpha'\,\beta\,R}\right)^4\,\sum_{N=1}^{\infty}\,
\frac{\e^{-\frac{\beta\,
      N}{\sqrt{2}\,R}}}{N}~\sum_{\substack{r\,z=N\\ r
    \,\textrm{odd}}}\,\frac1{r^4}~\sum_{\substack{x,y\in\zed/z\,\zed\\
y\neq0}}~\sum_{i=0}^9~
\int\limits_{\Delta}\frac{\dd^2\tau}{(\tau_2)^4}~\frac{\bigl(\,
\overline{\Theta}{}_i\bigr)^{14}}{\prod\limits_{j\neq i}\,
\overline{\Theta}{}_j}\\ && \times\,
\left[4\,\left(\frac{\Theta_7\,\Theta_8\,\Theta_9\,\Theta_0}
{\Theta_1\,\Theta_2\,\Theta_3\,\Theta_4\,\Theta_5\,\Theta_6}\right)^2
+\left(\frac{\Theta_2\,\Theta_3\,\Theta_5\,\Theta_7}
{\Theta_1\,\Theta_4\,\Theta_6\,\Theta_8\,\Theta_9\,\Theta_0}\right)^2
    -\left(\frac{\Theta_1\,\Theta_4\,\Theta_6\,\Theta_7}
{\Theta_2\,\Theta_3\,\Theta_5\,\Theta_8\,\Theta_9\,\Theta_0}\right)^2
\right.\nonumber\\ &&
+\,\left(\frac{\Theta_2\,\Theta_3\,\Theta_6\,\Theta_8}
{\Theta_1\,\Theta_4\,\Theta_5\,\Theta_7\,\Theta_9\,\Theta_0}\right)^2
-\left(\frac{\Theta_1\,\Theta_4\,\Theta_5\,\Theta_8}
{\Theta_2\,\Theta_3\,\Theta_6\,\Theta_7\,\Theta_9\,\Theta_0}\right)^2
    +\left(\frac{\Theta_3\,\Theta_4\,\Theta_5\,\Theta_9}
{\Theta_1\,\Theta_2\,\Theta_6\,\Theta_7\,\Theta_8\,\Theta_0}\right)^2
\nonumber\\ &&
-\left.\,\left(\frac{\Theta_1\,\Theta_2\,\Theta_6\,\Theta_9}
{\Theta_3\,\Theta_4\,\Theta_5\,\Theta_7\,\Theta_8\,\Theta_0}\right)^2
+\left(\frac{\Theta_3\,\Theta_4\,\Theta_6\,\Theta_0}
{\Theta_1\,\Theta_2\,\Theta_5\,\Theta_7\,\Theta_8\,\Theta_9}\right)^2
    -\left(\frac{\Theta_1\,\Theta_2\,\Theta_5\,\Theta_0}
{\Theta_3\,\Theta_4\,\Theta_6\,\Theta_7\,\Theta_8\,\Theta_9}\right)^2
\right] \ , \nonumber
\label{hetexptheta}\eea
which can again be expressed in terms of elliptic Jacobi-Erderlyi
functions by using the formula~(\ref{JacErd2}).

In the free string limit $g_s\to0$ the space of physical states of the
heterotic sigma-model on the symmetric product orbifold
(\ref{hetsymorb}) is naturally isomorphic to the Fock space of second
quantized heterotic strings in DLCQ~\cite{Rey1,Lowe1}. The
$(\zed_2)^N$ factor in this quotient space is a discrete gauge
symmetry acting on twisted sector gauge fermions $\lambda^A$ in the
fundamental representation of $G$. The additional $\zed_2$-orbifolds
are achieved by extra GSO projections on $\lambda^A$, and they are
necessary to reproduce the light-cone Green-Schwarz heterotic string field
theory~\cite{Rey1,Lowe1}. The right-moving sector is thus given by the
standard $\real^{24}/\zed_2$ orbifold conformal field theory. This
$\zed_2$-orbifold for $g_s>0$ is manifested through the decomposition
of the theta-constants comprising the modular form (\ref{Psi8def})
according to (\ref{JacErd2}), and it can be thought of as being
ultimately responsible in this instance for the fibred decomposition
of the Jacobian variety (\ref{extracommdiag}). Let us also remark that
in order to implement S-duality with Type~IB superstring theory (as is
necessary in formulating the heterotic matrix string theory
conjecture), one should add a Wilson line which breaks the heterotic
gauge group $G$ to $SO(16)\times SO(16)$~\cite{Rey1,Lowe1}. This may
be achieved by adding an appropriate $B$-field term
$\lambda^A\,B_{AB}\,\lambda^B$ to the heterotic string action
(\ref{heteroticaction}), whose effect is to simply modify the modular
form (\ref{Psi8def}) in a standard way. It amounts to a shift of the
imaginary part $\Omega_2$ of the period matrix of $\Sigma_2$ and thus
produces a reduction onto different tori in the right-moving
sector. This (non-modular) change of the base tori can be derived
directly from the corresponding Polyakov path integral~\cite{GOS1}.

\newsection{Boundary Contributions\label{DegLimits}}

In this final section we will elucidate some arithmetic and physical
aspects of the two-loop superstring free energy (\ref{F2exptheta}). We
have seen that the pertinent genus two theta-functions (\ref{JacErd2})
factorize into elliptic Jacobi-Erderlyi functions associated with the
fibration (\ref{extracommdiag}) of the Jacobian variety of the
original curve $\Sigma_2$ into two tori $\torus^2_{\ii\nu}$ and
$\torus^2_\tau$. But the resulting formulas for the free energies are
quite involved and difficult to deal with analytically. We will now
explore some regions of the moduli space ${\cal M}_2$ wherein this
factorization simplifies drastically and some precise information can
be extracted from these expressions.

\subsection{Pinching Parameters}

Let us begin with some general aspects concerning
the generic relationship between genus two curves and elliptic
curves. Generally, any genus two surface $\Sigma_2$ is a connected sum
$\Sigma_2=\torus^2_{\tau_1}\,\#\,\torus^2_{\tau_2}$ of two tori whose periods
can be expressed in terms of the moduli $\tau_i$ of the tori and a
complex number $ t $. The positive number $| t |<1$ is the
radius of the disks that are excised from the two tori in order to
sew them together to produce $\Sigma_2$. Let
$q_i:=\e^{2\pi\ii\tau_i}$, $i=1,2$. Then the pinching parameters
$q_1,q_2, t $ form an alternative set of moduli for
$\Sigma_2$.

The genus two period matrix $\Omega$ may be computed as a
holomorphic function of the pinching parameters
$q_1,q_2, t $~\cite{Yamada1}. For this, we use the sewing
formalism to express the holomorphic one-differential $\mbf\omega$ of
$\Sigma_2$ as a power series in $ t $ with coefficients
calculated from the genus one differentials $\omega^{(i)}$ of
$\torus^2_{\tau_i}$, $i=1,2$, and then use (\ref{periodnorms}) to
calculate the period matrix elements. We will need these
expressions only to leading order in $ t \to0$, in which case the
period matrix is given by
\bea
\Omega_{11}&=&\tau_1+\frac{ t ^2}{2\pi\ii}\,{\hat E}_2(q_1)+
O\left( t ^4\right) \ , \nonumber\\
\Omega_{22}&=&\tau_2+\frac{ t ^2}{2\pi\ii}\,{\hat E}_2(q_2)+
O\left( t ^4\right) \ , \nonumber\\
\Omega_{12}&=&-\frac{ t }{2\pi
  \ii}\,\Bigl(1+{\hat E_2}(q_1)\,{\hat E}_2(q_2)\, t ^2\Bigr)+
O\left( t ^5\right)
\label{Omegaepsilon}\eea
where
\beq
\hat E_2(q)=-\frac1{12}+2\,\sum_{n=1}^\infty\,\sigma_1(n)~q^n
\label{E2qdef}\eeq
is the normalized elliptic Eisenstein series, with $\sigma_1(n)$ the
number of $n$-sheeted unbranched covers of a torus given by
(\ref{sigma1Ndef}). Let us now specialize to the case where
$\Sigma_2\to\torus_{\ii\nu}^2$ is a branched covering with the reduced
form (\ref{simpleperiod}) of its period matrix. The moduli of the two
connecting tori can then be identified as $\tau_1=(x+\ii\frac r\nu)/z$
and $\tau_2=-\Omega_{22}=\tau$. The torus $\torus_{\tau_1}^2$ in this
case is an unbranched cover of the base $\torus_{\ii\nu}^2$ of degree
$N=r\,z$. The radius of the connecting cylinder may be identified as
$| t |=\frac yz$, which satisfies $0<| t |<1$ since
$y\in\zed/z\,\zed$ and $y\neq0$.

There are two classes of degenerations of the Riemann surface
$\Sigma_2$ up to modular transformations. When $ t \to0$, the
connecting cylinder is pinched down and $\Sigma_2$ degenerates into
the two tori $\torus_{\tau_1}^2$ and $\torus_{\tau_2}^2$. This
provides a geometric description of the moduli space ${\cal M}_2$ near
the divisor of surfaces $\Sigma_2$ with nodes, and it corresponds to
the limit in which the two branch points on $\torus_{\ii\nu}^2$
coincide (singularity type~(b) in the terminology of
Section~\ref{2LoopWorld}). When $q_i\to0$ for $i=1$ or $i=2$,
i.e. $\tau_i\to\ii\infty$, the torus $\torus^2_{\tau_i}$ degenerates
to a Riemann sphere by making its homology cycle $\beta$ infinitely
long, or equivalently by modular invariance shrinking the cycle to
zero size (singularity type~(c)). It is straightforward to see that
the other boundary limits of
the moduli space ${\cal M}_2$, determined by the positivity condition
(\ref{Omega2pos}), can be mapped into these other two cases. Let us
now examine each of these limits in some detail.

\subsection{Factorization\label{Factorization}}

In the sewing construction one may view the genus two surface
$\Sigma_2$ as the disjoint union
$\Sigma_2=\elliptic\,_1\amalg\ann_{\,t}\amalg\elliptic\,_2$, where
$\ann_{\,t}=\{(z_1,z_2)\in\complex^2~|~z_i\in\disk^2 \ , \ z_1\,z_2=t\}$
for $t\neq0$ is the annulus with outer radius~$1$ and inner radius
$t$, $\disk^2$ is the unit disk in $\complex$, and
$\elliptic\,_i=\torus_{\tau_i}^2\setminus\disk^2$ with $z_i$ local
complex coordinates on $\torus_{\tau_i}^2$. In conformal field theory,
the surfaces with boundary $\elliptic\,_i$, $i=1,2$ define two states
$\langle\elliptic\,_1|$ and $|\elliptic\,_2\rangle$. Within the Hamiltonian
framework, we identify the annulus with a cylinder via the exponential
map. The cylinder amplitude then corresponds to the operator insertion
$t^{L_0}\,\overline{t}{}^{\,\overline{L}{}_0}$, where
$L_0+\overline{L}{}_0$ is the worldsheet Hamiltonian and
$L_0-\overline{L}{}_0$ is the momentum operator.

The genus two superstring free energy is then given symbolically by
\beq
F_2=\bigl\langle\elliptic\,_1\bigl|\,t^{L_0}\,\overline{t}{}^{\,
\overline{L}{}_0}\,\bigr|\elliptic\,_2\bigr\rangle \ .
\label{F2CFTdef}\eeq
We can insert a complete set of states into the matrix element
(\ref{F2CFTdef}) which diagonalize the Virasoro operators
$L_0,\overline{L}{}_0$ to get
\beq
F_2=\sum_I\,\bigl\langle\elliptic\,_1\big|\psi_I\bigr\rangle\,
\bigl\langle\psi_I\bigl|\,t^{L_0}\,\overline{t}{}^{\,
\overline{L}{}_0}\,\bigr|\psi_I\bigr\rangle\,\bigl\langle
\psi_I\big|\elliptic\,_2\bigr\rangle \ .
\label{complsetinsert}\eeq
This yields a Laurent series expansion in
$|t|$. After GSO projection, the leading contribution comes from the
massless vacuum states having $L_0=\overline{L}{}_0=0$ and zero
momentum, so that
\beq
F_2=F^{(1)}\,F^{(2)}+O\bigl(|t|\bigr) \ ,
\label{F2leading}\eeq
where $F^{(i)}$ is the one-loop free energy for the torus
$\torus^2_{\tau_i}$. We should stress that the expression
(\ref{F2leading}) is only meant to be symbolic. In particular, it is
only valid at fixed spin structure and fixed winding
numbers around the finite temperature DLCQ torus
$\torus_{\ii\nu}^2$, in which case the leading term is actually down
by a negative power of $|t|$. Summing over these quantum numbers mixes
the two one-loop contributions in a non-trivial way and spoils the
explicit factorization of the leading order term. We shall see this
explicitly below. The higher-order terms in (\ref{F2leading}) arise
from propagation of massless physical states in the long thin tube
connecting the two tori~\cite{AtMS1}, and in this limit the genus two
free energy is related to a sum of products of one-loop tadpoles for
the massless states represented as torus one-point functions.

We will now identify these one-loop string theories. Let
$\delta=\big({{\mbf a_1}\atop{\mbf a_2}}\big)\neq\delta_0$ be any even
genus two spin structure such that $\mbf a_i\in\{0,1\}^2$ is an even
genus one spin structure on $\torus_{\tau_i}^2$. In the limit $t\to0$,
the leading asymptotics of the genus two theta-constants are given by
\bea
\Theta[\delta](\mbf0|\Omega)&=&\theta[\mbf a_1]
(0|\tau_1)\,\theta[\mbf a_2](0|\tau_2)+O\left(t^2\right) \ ,
\nonumber\\ \Theta[\delta_0](\mbf 0|\Omega)&=&t~
\eta(\tau_1)^3\,\eta(\tau_2)^3+O\left(t^3\right) \ ,
\label{Theta2lead}\eea
which implies that the cusp form (\ref{chi10}) has the leading
asymptotics
\beq
\Psi_{10}(\Omega)=t^2~\eta(\tau_1)^{24}\,
\eta(\tau_2)^{24}+O\left(t^4\right) \ .
\label{chi10asympt}\eeq
It is instructive to first examine the behaviour of the bosonic free
energy (\ref{bosfreefinal}) in this limit. Notice, first of all, that
since $y\in\zed/z\,\zed$ with $y\neq0$, the limit $t\to0$ is
equivalent to taking $z\to\infty$, i.e. the limit $N\to\infty$ of
branched covers with large degree. This means that we should look at
the large $N$ asymptotic tail behaviour of the series
(\ref{bosfreefinal}). One then has
\bea
\lim_{z\to\infty}\,F_2^{\rm bos}&=&-g_s^2\,\left(\frac1
{4\,\sqrt2\,\pi\,\beta\,R}\right)^{12}\,
\sum_{N=1}^\infty\,\e^{-\frac{\beta\,N}{\sqrt2\,R}}~
\sum_{r\,z=N}\,\left(\frac zr\right)^{12}\nn\\ && \times\,
\sum_{\substack{x,y\in\zed/z\,\zed\\ y\neq0}}\,
\frac1{y^4}\,{\cal Z}_1^{\rm bos}\bigl(\tau'\,,\,\overline{
\tau}{\,}'\,\bigr)\Big|_{\tau'=\frac{x+\frac{\ii r}\nu}z}~\tilde
F_1^{\rm bos}
\label{F2bosasympt}\eea
where
\beq
\tilde F_1^{\rm bos}=\int\limits_\Delta\,
\frac{\dd^2\tau}{(\tau_2)^{12}}~{\cal Z}_1^{\rm bos}(\tau,
\overline{\tau}\,)
\label{tildeF1bos}\eeq
and ${\cal Z}_1^{\rm bos}(\tau,\overline{\tau}\,)=\Tr\,q^{L_0-2}\,
\overline{q}{}^{\,\overline{L}{}_0-2}=|\eta(\tau\,)|^{-48}$ is the
one-loop first quantized partition function on
$\torus_{\tau}^2$. Thus the contribution of the unramified coverings
of $\torus_{\ii\nu}^2$ is the same as in the one-loop computation of
Section~\ref{OneLoop}, while the contribution over the auxilliary
torus $\torus^2_{\tau}$ resembles the second quantized one-loop
bosonic partition function (Note that this is {\it not} the standard
$SL(2,\zed)$ modular invariant partition function, as modular
invariance of the expression (\ref{F2bosasympt}) under the genus two
residual modular group ${\cal G}\subset Sp(4,\zed)$ is required
here). This is a twisted admixture of the operation providing the
mapping from first quantization to second quantization that was given
by Hecke transforms in Section~\ref{OneLoop}.

To understand the algebraic meaning of the mapping in the present
case, we now turn our attention to the superstring free energy
(\ref{F2exptheta}). For this, we also need the asymptotic behaviours
of the quantities (\ref{Xi6def}), which from (\ref{Theta2lead}) and
the formulas of Appendix~B can be computed to be
\bea
\Xi_6[\delta](\Omega)&=&-2^8\,\bigl\langle\mbf a_1\big|\big({}^1_1\big)
\bigr\rangle\,\bigl\langle\mbf a_2\big|\big({}^1_1\big)
\bigr\rangle~\eta(\tau_1)^{12}\,\eta(\tau_2)^{12}+O\left(t^2\right) \
, \nonumber\\\Xi_6[\delta_0](\Omega)&=&-3\cdot2^8\,~\eta(\tau_1)^{12}\,
\eta(\tau_2)^{12}+O\left(t^2\right) \ .
\label{Xi6asympt}\eea
Substituting (\ref{Theta2lead}) and (\ref{Xi6asympt}) into the
numerator of the integrand in (\ref{finalfreeenergy}), one finds that
the contributions from the spin structures $\delta_7$, $\delta_8$ and
$\delta_9$ sum to~$0$ by the Jacobi abstruse identity
(\ref{abstruse}). This sum is tantamount to a partial GSO
projection which removes the would be tachyonic divergence coming from
(\ref{chi10asympt}) in the degeneration limit $t\to0$. Only the
contribution from the spin structure $\delta_0$ remains, and
(\ref{finalfreeenergy}) becomes
\bea
\lim_{z\to\infty}\,F_2&=&-\frac{g_s^2}{4}\,\left(\frac1
{4\,\sqrt2\,\pi\,\beta\,R}\right)^4\,
\sum_{N=1}^\infty\,\frac{\e^{-\frac{\beta\,N}{\sqrt2\,R}}}N~
\sum_{\substack{r\,z=N\\r\,{\rm odd}}}\,\frac1{r^4}
{}~\sum_{\substack{x,y\in\zed/z\,\zed\\ y\neq0}}~
\int\limits_\Delta\,\frac{\dd^2\tau}{(\tau_2)^4}~\left(
\frac{3\pi^2}4\,\frac{y^2}{z^2}\right)^2\nonumber\\ &=&-
\frac{\sqrt3\,\pi^2\,g_s^2}{8}\,\left(\frac1
{4\,\sqrt2\,\pi\,\beta\,R}\right)^4\,
\sum_{N=1}^\infty\,\frac{\e^{-\frac{\beta\,N}{\sqrt2\,R}}}N~
\sum_{\substack{r\,z=N\\r\,{\rm odd}}}\,\frac1{r^4}\,\left(
\mbox{$\frac15\,z^2-\frac12\,z$}\right)
\label{F2asymptgen}\eea
to $O(z^{-1})$. The removal of the tachyonic divergence from $\tilde
F_2^{\rm bos}$ in (\ref{F2bosasympt}) has completely trivialized the
partition function over the auxilliary torus and the only contribution
that remains is from the unbranched cover over
$\torus^2_{\ii\nu}$. The precise form of this sum is now determined by
the way in which we analyse the large degree asymptotics as
$N\to\infty$ of this series.

Let us first take the limit $z\to\infty$ with $r$ finite. In this limit
$\tau_1\to0$ and the covering torus $\Sigma_1$ shrinks to a
point. Nevertheless, some remnant of the genus two covering map
remains due to the fibration over the auxilliary torus $\torus_\tau^2$
in (\ref{extracommdiag}). In this regime we may disregard the odd
parity constraint on the sum over the divisors $r$ in
(\ref{F2asymptgen}), and the free energy thereby becomes
\beq
\lim_{\substack{z\to\infty\\ r\ll
    z}}\,F_2=-\frac{\sqrt3\,\pi^2\,g_s^2}
{8}\,\left(\frac1{4\,\sqrt2\,\pi\,\beta\,R}\right)^4\,
\sum_{N=1}^\infty\,\frac{\e^{-\frac{\beta\,N}
{\sqrt2\,R}}}{N^5}~\left(\mbox{$\frac15\,\sigma_6(N)-\frac12\,
\sigma_5(N)$}\right)
\label{F2asymptrllz}\eeq
where the divisor functions
\beq
\sigma_k(N)=\sum_{z|N}\,z^k
\label{sigmakNdef}\eeq
generalize the integers $\sigma_1(N)$ in (\ref{sigma1Ndef}) which
count the unramified coverings $\Sigma_1$ of $\torus_{\ii\nu}^2$. The
series (\ref{F2asymptrllz}) can be naturally related to the Hecke
algebra as follows.

Consider the lattice $\Lambda_\tau:=\zed\oplus\zed\,\tau$ such that
$\torus^2_\tau=\complex/\Lambda_\tau$. For any integer $k\geq2$,
introduce the holomorphic Eisenstein series~\cite{Apostol1}
\beq
G_{2k}(\tau):=\sum_{\substack{\lambda\in\Lambda_\tau\\\lambda\neq(0,0)}}
\,\frac1{\lambda^{2k}}=
2\,\zeta(2k)+2\,\frac{(2\pi\ii)^k}{(2k-1)!}\,\sum_{n=1}^\infty\,
\sigma_{2k-1}(n)~q^n
\label{Gktaudef}\eeq
with $q=\e^{2\pi\ii\tau}$. This defines a modular form of weight
$2k$. The action on (\ref{Gktaudef}) of the Hecke operator ${\bf H}_N$
defined in (\ref{Heckeop}) is given by
\beq
{\bf
  H}_N*G_{2k}(\tau)=N^{2k-1}\,\sum_{\substack{\Lambda'\subset\Lambda_\tau
\\ [\Lambda_\tau:\Lambda'\,]=N}}~\sum_{\substack{\lambda\in\Lambda'\\
\lambda\neq(0,0)}}\,\frac1{\lambda^{2k}} \ .
\label{HNG2k1}\eeq
To work out this sum explicitly, suppose first that $N=p$ is a prime
number. If $\lambda\in p\,\Lambda_\tau$, then $\lambda$
lies in all sublattices $\Lambda'$ of $\Lambda_\tau$ of index~$p$ and
so contributes
$\frac{\sigma_1(p)}{\lambda^{2k}}=\frac{p+1}{\lambda^{2k}}$ to the sum
(\ref{HNG2k1}). Otherwise, $\lambda$ lies in only one sublattice
$\Lambda'=p\,\Lambda_\tau\oplus\zed\,\lambda$ and so contributes
$\frac1{\lambda^{2k}}$. Thus
\beq
{\bf H}_p*G_{2k}(\tau)=p^{2k-1}~G_{2k}(\tau)+p^{2k}~
\sum_{\substack{\lambda\in p\,\Lambda_\tau \\ \lambda\neq(0,0)}}\,
\frac1{\lambda^{2k}}=p^{2k-1}~G_{2k}(\tau)+G_{2k}(\tau)
\label{HpG2ktau}\eeq
and it follows that $G_{2k}(\tau)$ is an eigenform of ${\bf H}_p$ with
eigenvalue $\sigma_{2k-1}(p)=1+p^{2k-1}$. In the general case, we
use the prime factorization of the integer $N$ along with the Hecke
algebra property ${\bf H}_n\circ{\bf H}_m={\bf H}_{n\,m}$ for ${\rm
  gcd}(n,m)=1$ to conclude that the Eisenstein series $G_{2k}$ is a
simultaneous eigenform of each Hecke operator ${\bf H}_N$ with
eigenvalue $\sigma_{2k-1}(N)$. Similarly, each ${\bf H}_N$ has
eigenforms comprised of elliptic cusp forms
$\eta(\tau)^{24}\,G_{2k}(\tau)$~\cite{Apostol1}.

Let us now take the limit $z\to\infty$ with $r\sim z$. In this limit
$\tau_1\to\frac\ii\nu$ and the surface $\Sigma_2$ factorizes into the
original spacetime torus $\torus_{\ii\nu}^2$ (up to a modular
transformation) and the auxilliary torus $\torus_\tau^2$. The free
energy (\ref{F2asymptgen}) in this regime vanishes,
\beq
\lim_{\substack{z\to\infty\\ r\sim z}}\,F_2=0 \ ,
\label{F2asymptrsimz}\eeq
to leading order. At this order supersymmetry is restored by the
factorization and there are no contributions from this boundary
component of the moduli space ${\cal M}_2$. The combinatorics of the
covers in these factorizing degeneration limits are thereby accounted
for by a sort of ``topological'' string theory which counts particular
eigenvalues in the spectra of the Hecke operators. The role of the
degenerate free energy as a generating function for the Hecke spectra will also
persist at higher orders in the cylindrical length $t$. For example,
the Siegel cusp form of weight ten has has the leading
expansion~\cite{Tuite1}
\beq
\Psi_{10}(\Omega)^{-1}=t^{-2}\,\eta(\tau_1)^{-24}\,
\eta(\tau_2)^{-24}\,\left[1+12\,t^2\,\hat E(q_1)\,\hat E(q_2)+
O\left(t^4\right)\right]
\label{Psi10t2}\eeq
as $t\to0$.

\subsection{Collapsing Homology Cycles\label{Handle}}

Let us now look at the limit $q_1\to0$ in which the handle with
homology cycles $a_1,b_1$ degenerates. In this non-separating
degeneration limit, the surface $\Sigma_2$ becomes the auxilliary
torus $\torus_{\tau}^2$. If $\mbf a\in\{0,1\}^2$ is any even genus one
characteristic, then the even characteristic genus two theta-constants
generally have a power series expansion around $q_1=0$ given by
\bea
\Theta\bigl[{}^{\,\mbf a}_{00}\bigr](\mbf 0|\Omega)&=&
\sum_{n=-\infty}^\infty\,(q_1)^{n^2}~\theta[\mbf a]\bigl(-
\mbox{$\frac{n\,t}{2\pi\ii}$}\big|\tau_2\bigr) \ , \nonumber\\
\Theta\bigl[{}^{\,\mbf a}_{01}\bigr](\mbf 0|\Omega)&=&
\sum_{n=-\infty}^\infty\,(-1)^n\,(q_1)^{n^2}~\theta[\mbf a]\bigl(-
\mbox{$\frac{n\,t}{2\pi\ii}$}\big|\tau_2\bigr) \ , \nonumber\\
\Theta\bigl[{}^{\,\mbf a}_{10}\bigr](\mbf 0|\Omega)&=&
\sum_{n=-\infty}^\infty\,(q_1)^{(n+\frac12)^2}~\theta[\mbf a]\bigl(-
\mbox{$\frac{(n+\frac12)\,t}{2\pi\ii}$}\big|\tau_2\bigr) \ , \nonumber\\
\Theta\bigl[\delta_0\bigr](\mbf 0|\Omega)&=&
\sum_{n=-\infty}^\infty\,\ii(-1)^n\,(q_1)^{(n+\frac12)^2}~\theta_1\bigl(-
\mbox{$\frac{(n+\frac12)\,t}{2\pi\ii}$}\big|\tau_2\bigr) \ .
\label{Thetaq0asympt}\eea
It follows that the cusp form (\ref{chi10}) has the leading
asymptotics
\beq
\Psi_{10}(\Omega)=-(q_1)^2\,\eta(\tau_2)^{18}\,\theta_1\bigl(
-\mbox{$\frac t{4\pi\ii}$}\big|\tau_2\bigr)^2+O\left((q_1)^2\right) \ .
\label{Psi10q0asympt}\eeq
After some algebra using the formulas of Appendix~B, one thereby finds
that leading behaviour of the free energy (\ref{finalfreeenergy}) is
given by
\bea
\lim_{q_1\to0}\,F_2&=&-\frac{g_s^2}{32}\,\left(\frac1
{4\,\sqrt2\,\pi\,\beta\,R}\right)^4\,\sum_{N=1}^{\infty}\,
\e^{-\frac{\beta\,N}{\sqrt{2}\,R}}~\sum_{\substack{r\,z=N\\ r
    \,\textrm{odd}}}\,\frac1{r^5}~\sum_{\substack{y\in\zed/z\,\zed\\
y\neq0}}~\int\limits_{\Delta}
\frac{\dd^2\tau}{(\tau_2)^4}~\frac1{\Big|\eta(\tau)\Big|^{36}\,
\Big|\theta_1\bigl(
\mbox{$\frac y{2\,z}$}\big|\tau\bigr)\Big|^4}\nonumber\\ && \times\,\left|
\theta_4(0|\tau)^8\,\left[\theta_4\bigl(\mbox{$\frac y{2\,z}$}\big|
\tau\bigr)^4\,\theta_1\bigl(\mbox{$\frac y{2\,z}$}\big|
\tau\bigr)^4+\theta_2\bigl(\mbox{$\frac y{2\,z}$}\big|
\tau\bigr)^4\,\theta_3\bigl(\mbox{$\frac y{2\,z}$}\big|
\tau\bigr)^4\right]\right.\nonumber\\ && +~
\theta_2(0|\tau)^8\,\left[\theta_4\bigl(\mbox{$\frac y{2\,z}$}\big|
\tau\bigr)^4\,\theta_3\bigl(\mbox{$\frac y{2\,z}$}\big|
\tau\bigr)^4+\theta_2\bigl(\mbox{$\frac y{2\,z}$}\big|
\tau\bigr)^4\,\theta_1\bigl(\mbox{$\frac y{2\,z}$}\big|
\tau\bigr)^4\right]\nonumber\\ &&-\left.\,
\theta_3(0|\tau)^8\,\left[\theta_1\bigl(\mbox{$\frac y{2\,z}$}\big|
\tau\bigr)^4\,\theta_3\bigl(\mbox{$\frac y{2\,z}$}\big|
\tau\bigr)^4+\theta_2\bigl(\mbox{$\frac y{2\,z}$}\big|
\tau\bigr)^4\,\theta_4\bigl(\mbox{$\frac y{2\,z}$}\big|
\tau\bigr)^4\right]\right|^2 \ .
\label{F2q1to0}\eea
The elliptic modular integrals in (\ref{F2q1to0}) are finite.

The degeneration limit $q_1\to0$ corresponds to the shrinking limit
$\nu\rightarrow 0$ of the original spacetime torus
$\torus^2_{\ii\nu}$. There are two ways in which we can make the
parameter (\ref{nudef}) vanish. Taking $\beta\to\infty$ gives the zero
temperature limit of the free energy, which is proportional to the
vacuum energy. Since $N\geq1$, all terms in the series are
exponentially damped and thus the vacuum energy vanishes, as expected
since this limit simply corresponds to the restoration of
supersymmetry at zero temperature. On the other hand, taking
$R\to\infty$ decompactifies the light cone and sends the exponential
factors to~$1$ in (\ref{F2q1to0}). Apart from an overall factor, the
free energy is then independent of temperature, except for its
dependence on the winding number $r$. In this case the strings
effectively propagate on a $\zed_2$ orbifold of flat
space~\cite{ADHPh1} defined by the antiperiodic fermion boundary
conditions, which is presumably a subsector of the symmetric orbifold
superconformal field theory on $\real^8$ for each $N$. This string
theory is non-supersymmetric and hence has a non-vanishing vacuum
energy corresponding to contributions from physical
tachyons~\cite{AtMS1}. In each of these decompactification limits, the
discrete data of the branched cover should assemble themselves into a
continuum limit which restores the two complex dimensions of the
moduli space ${\cal M}_2$~\cite{BBNT1}.

Let us now consider the non-separating degeneration limit $q_2\to0$ in
which the branched cover $\Sigma_2$ becomes an unramified covering of
the original spacetime torus $\torus_{\ii\nu}^2$ (up to a modular
transformation). This corresponds to the contributions from the
$\tau\to\ii\infty$ region of the elliptic modular integral in
(\ref{finalfreeenergy}). We may use the same asymptotic formulas
(\ref{Thetaq0asympt}) and (\ref{Psi10q0asympt}) with $q_1,\tau_2$
replaced by $q_2,\tau_1$. The terms
$\Xi_6[\delta_i]\,\Theta[\delta_i](\mbf 0|\Omega)^4$ for $i=7,8$ have
leading terms of order $q_2$. These two terms thus give a contribution
to the integration over moduli space which has a simple pole at
$q_2=0$. This divergence arises from the tachyon traversing the
$a_2$ cycle of the elliptic component $\torus_\tau^2$ of the
degeneration~\cite{AtMS1,DHPh1}. However, the sum
$\Xi_6[\delta_7]\,\Theta[\delta_7](\mbf
0|\Omega)^4+\Xi_6[\delta_8]\,\Theta[\delta_8](\mbf 0|\Omega)^4$ is
found to vanish to this order and thus removes the pole. This
corresponds to a partial GSO projection in the Neveu-Schwarz sector of
the genus one component $\torus_\tau^2$ which eliminates the
tachyon. The contributions from the remaining spin structures
$\delta_0$ and $\delta_9$ correspond to Ramond states propagating in
$\torus_\tau^2$ and are of order $(q_2)^2$, yielding no poles.

Working out each of the four contributions to (\ref{finalfreeenergy})
up to order $(q_2)^2$ leads after some algebra to the free energy
\beq
\lim_{q_2\to0}\,F_2=-\frac{g_s^2}{64}\,\left(\frac1
{4\,\sqrt2\,\pi\,\beta\,R}\right)^4\,
\sum_{N=1}^{\infty}\,\frac{\e^{-\frac{\beta\,
      N}{\sqrt{2}\,R}}}{N}~\sum_{\substack{r\,z=N\\ r
    \,\textrm{odd}}}\,\frac1{r^4}~\sum_{\substack{x,y\in\zed/z\,\zed\\
y\neq0}}\,\Bigl|{\cal Z}_1^\infty(\zeta|\tau_1)\Bigr|^2
\Bigg|_{\substack{\zeta=\frac yz\\ \tau_1=\frac{x+\ii\frac r\nu}z}} \
,
\label{F2q2to0}\eeq
where
\bea
{\cal
  Z}_1^\infty(\zeta|\tau_1)&=&\frac1{\eta(\tau_1)^{18}\,\theta_1
\bigl(\mbox{$\frac\zeta2$}\big|
\tau_1\bigr)^2}\,\Biggl[2\,\theta_2(0|\tau_1)^8\,\Bigl(\theta_1\bigl(
\mbox{$\frac\zeta2$}\big|\tau_1\bigr)^4\,\theta_2\bigl(
\mbox{$\frac\zeta2$}\big|\tau_1\bigr)^4+\theta_3\bigl(
\mbox{$\frac\zeta2$}\big|\tau_1\bigr)^4\,\theta_4\bigl(
\mbox{$\frac\zeta2$}\big|\tau_1\bigr)^4\Bigr)\Biggr.\nonumber\\
&& -\,\theta_3(0|\tau_1)^8\,\Bigl(\theta_1\bigl(
\mbox{$\frac\zeta2$}\big|\tau_1\bigr)^4\,\theta_3\bigl(
\mbox{$\frac\zeta2$}\big|\tau_1\bigr)^4+\theta_2\bigl(
\mbox{$\frac\zeta2$}\big|\tau_1\bigr)^4\,\theta_4\bigl(
\mbox{$\frac\zeta2$}\big|\tau_1\bigr)^4\Bigr)\nonumber\\
&& +\,\theta_4(0|\tau_1)^8\,\Bigl(\theta_1\bigl(
\mbox{$\frac\zeta2$}\big|\tau_1\bigr)^4\,\theta_4\bigl(
\mbox{$\frac\zeta2$}\big|\tau_1\bigr)^4+\theta_2\bigl(
\mbox{$\frac\zeta2$}\big|\tau_1\bigr)^4\,\theta_3\bigl(
\mbox{$\frac\zeta2$}\big|\tau_1\bigr)^4\Bigr)\nonumber\\
&&-\,8\,\eta(\tau_1)^3\,\theta_1\bigl(
\mbox{$\frac\zeta2$}\big|\tau_1\bigr)^4\,\Bigl(\theta_2(0|\tau_1)\,
\theta_3(0|\tau_1)\,\theta_4(\zeta|\tau_1)+\theta_2(0|\tau_1)\,
\theta_4(0|\tau_1)\,\theta_3(\zeta|\tau_1)\Bigr.\nonumber\\
&&\qquad\qquad\qquad\qquad\qquad\qquad+\Bigl.\,\theta_3(0|\tau_1)\,
\theta_4(0|\tau_1)\,\theta_2(\zeta|\tau_1)\Bigr)\nonumber\\
&&-\,8\,\eta(\tau_1)^3\,\theta_2\bigl(
\mbox{$\frac\zeta2$}\big|\tau_1\bigr)^4\,\Bigl(\theta_2(0|\tau_1)\,
\theta_3(0|\tau_1)\,\theta_4(\zeta|\tau_1)+\theta_2(0|\tau_1)\,
\theta_4(0|\tau_1)\,\theta_3(\zeta|\tau_1)\Bigr.\nonumber\\
&&\qquad\qquad\qquad\qquad\qquad\qquad-\Biggl.\Bigl.\,\theta_3(0|\tau_1)\,
\theta_4(0|\tau_1)\,\theta_2(\zeta|\tau_1)\Bigr)\Biggr]
\label{calZ1infty}\eea
and we have dropped an irrelevant overall numerical constant in
(\ref{F2q2to0}) arising from the remaining modular integration over
$\tau_2\in\Delta$. As before, the non-vanishing of this boundary
contribution is due to the presence of physical tachyons. This free
energy is a natural extension of the one-loop result of
Section~\ref{OneLoop}, illustrating the appropriate modification for
the action of the Hecke algebra at two-loops.

\subsection*{Acknowledgments}

We thank H.~Braden, C.-S.~Chu, P.~Di~Vecchia, B.~Dolan,
J.C.~Eilbeck, V.~Enolski, J.~Howie, O.~Lechtenfeld, N.~Obers,
P.~Orland, A.~Morozov, S.~Ramgoolam and G.~Semenoff for helpful discussions and
correspondence. This work was supported in part by the EU-RTN Network
Grant MRTN-CT-2004-005104. The work of H.C. was supported in part by
an EPSRC Postgraduate Studentship. The work of R.J.S. was supported in
part by PPARC Grant PPA/G/S/2002/00478.

\setcounter{section}{0}
\setcounter{subsection}{0}
\setcounter{subsubsection}{0}

\appendix{Moduli Space for the Poincar\'e Normal Form\label{PNOcalc}}

In this appendix we will sketch the computation of the two-loop free
energy from the fully reduced Poincar\'e normal form (\ref{PNF}). This
is done for the sake of completeness and because it provides some
interesting alternative characterizations of the genus two Hurwitz
moduli space which may be of independent interest. As we will see, the
free energy in this case cannot be made as explicit as in the main
text, but the same reduction features do carry through nonetheless.

\subsection{Reduced Moduli}

The genus two Poincar\'e normal form is given by
\be {\sf P}=r\,\left(\begin{array}{cccc}1 & 0 & 0 &  0\\
    0 & s & t & 0\end{array}\right) \ . \label{PNFapp}\ee
The matrix $\sf T$ which appears in the Frobenius normal
form (\ref{Frobform}) can be absorbed into the period matrix as in
(\ref{Tabsorb}) but the symplectic unimodular matrix
\beq
{\sf S}=\left(\begin{array}{cc}a & b \\ c &
    d\end{array}\right) \ ,
\label{sfSdef}\eeq
which acts on the base torus $\torus^2_{\ii\nu}$ as a modular
transformation, remains~\cite{Martens1,Igusa1}. This is one of the
reasons why the full reduction
is undesirable, as both the moduli space and the GSO projection
depend explicitly on the four integers $a,b,c,d$ which are functions
of the parameters $r$, $s$ and $t$. We have to keep $\sf S$ explicitly
in all of our calculations, and then sum over all the corresponding
$SL(2,\zed)$ modular transformations of the base. On the other hand,
the full reduction to (\ref{PNFapp}) leads to a somewhat simpler
decomposition of genus two theta-constants into elliptic
theta-functions~\cite{BelEn1}.

Given the homology matrix (\ref{Frobform}), we can rewrite the constraint
equation (\ref{matrixconstraint}) using (\ref{PNFapp}) and (\ref{sfSdef})
to get
\begin{equation}
    \label{newmatrixconstraint}{\sf H}^\top\,
(\id_2,\Omega)=(1,\ii\nu)\,{\sf S}\,
{\sf P}\,{\sf T}=
    (1,\ii\nu)\,\left(\begin{array}{cccc} r\,a & r\,b\,s & r\,b\,t &  0\\
    r\,c & r\,d\,s & r\,d\,t & 0\end{array}\right)\, {\sf T} \ .
\end{equation}
In order to factorize the genus two theta-constants in terms of
elliptic functions as in Section~\ref{ThetaFuns}, the period matrix
must have rational-valued off-diagonal elements. This will happen if we
modify (\ref{newmatrixconstraint}) by multiplying $\sf P$ with the
intersection form $-{\sf J}_{2}=({\sf J}_2)^{-1}$ to obtain
\begin{equation}\label{red}
    {\sf H}^\top\,(\id_2,\Omega)=
    (1,\ii\nu)\,\left(\begin{array}{cccc} r\,b\,t & 0 & -r\,a &-r\,b\,s\\
    r\,d\,t & 0 &-r\,c & -r\,d\,s\end{array}\right)\,{\sf J}_{2}\,{\sf
  T} \ .
\end{equation}
The matrix ${\sf J}_{2}\,{\sf T}\in Sp(4,\zed)$ is invertible. The
inverse $({\sf J}_{2}\,{\sf T})^{-1}$ acts on the left-hand side of
(\ref{red}) as a modular transformation on the period matrix $\Omega$
and on the pullback matrix ${\sf H}$ as in (\ref{Tabsorb}).

We can now solve the constraint equation for the period matrix by
first computing
\beq
{\sf H}=(1,\ii\nu)\,\left(\begin{array}{cc}r\,b\,t & 0
    \\ r\,d\,t
    & 0 \end{array}\right)=r\,t\,(b+d\ii\nu,0)
\eeq
to get
\beq
{\sf H}\,\Omega=r\,t\,(b+d\ii\nu,0)\,
\left(\begin{array}{cc}\Omega_{11} & \Omega_{12} \\ \Omega_{12} &
    \Omega_{22} \end{array}\right)=(1,\ii\nu)\,
\left(\begin{array}{cc}-r\,a & -r\,b\,s \\ -r\,c &
    -r\,d\,s \end{array}\right) \ .
\eeq
After a $\zed_2$ reflection, we thereby find
\be\label{nicepm}  \Omega=\left(\begin{array}{cc} \frac{\tau_\nu}{\mu\,s}
    & \frac{1}{\mu} \\ \frac{1}{\mu} &
    \tau \end{array}\right)\ee
where the integer $\mu$ is defined through $t=\mu\,s$ and is related
to the degree $N$ of the cover by $N=r^2\,\mu\,s$. As before
$\tau:=\Omega_{22}\in{\cal H}_1$ parametrizes an auxilliary torus
$\torus_\tau^2$, while
\beq
\tau_\nu=\frac{a+c\,(\ii\nu)}{b+d\,(\ii\nu)}
\eeq
labels a torus $\torus^2_{\tau_\nu}$ in the same elliptic modular orbit as the
base $\torus_{\ii\nu}^2$. This is in contrast to the partial reduction
carried out in the main text, in which the discretely parametrized
tori were unramified covers $\Sigma_1$ over the base.

\subsection{Residual Modular Group}

The Poincar\'e normal form is obtained through a change of homology
basis of $\Sigma_2$. The residual modular group ${\cal G}$ is the
subgroup of $Sp(4,\zed)$ which preserves the form (\ref{red}). It
consists of integral matrices of the form
\be\left(\begin{array}{cccc} 1 & -\mu\,\alpha & \alpha & \beta \\ 0 &
    1-\mu\,\gamma & \gamma & \delta \\ 0 & 0 & 1 &
    0 \\ 0 & -\mu^2\,\alpha & \mu\,\alpha &
    1+\mu\,\beta\end{array}\right)\ee
which obey the $Sp(4,\zed)$ condition
\be\gamma-\beta=\mu\,(\alpha\,\delta-\beta\,\gamma) \ . \ee
The extended fundamental domain ${\cal
  F}'_2={{\cal H}_2}/{{\cal G}}$ is then constructed as the quotient
of the Siegal upper half-plane by the residual modular group.

By using the $Sp(4,\zed)$ transformation rule (\ref{Omegamodular}),
one finds that under the action of the residual modular group the
period matrix elements transform as
\bea
\Omega_{22}&\longmapsto&\frac{\Omega_{22}\,(1-\mu\,\gamma)+\delta}{1+\mu\,
\beta-\mu^2\,\alpha\,\Omega_{22}} \ , \nonumber\\
\Omega_{12}&\longmapsto&\frac{\Omega_{12}-\mu\,\alpha\,\Omega_{22}+\beta}
{1+\mu\,\beta-\mu^2\,\alpha\,\Omega_{22}} \ , \nonumber\\
\Omega_{11}&\longmapsto&\Omega_{11}+\frac{\alpha\,(\mu\,\Omega_{12}-1)^2}
{1+\mu\,\beta-\mu^2\,\alpha\,\Omega_{22}} \ .
\label{sl2subgroup}\eea
The M\"obius transformations of $\tau=\Omega_{22}$ in the first line of
(\ref{sl2subgroup}) form a congruence subgroup of the elliptic modular
group $SL(2,\zed)$ defined by
\be
\Gamma_{(\mu)}=\Bigg\{\left(\begin{array}{cc}a & b \\ c &
    d\end{array}\right)\in SL(2,\zed)\quad\Bigg|\quad
a,d\equiv1~{\rm mod}\,\mu \ , \ c\equiv0~{\rm mod}\,\mu^2 \ ,
\ b\in\zed\Bigg\} \ . \ee
Once the elliptic fundamental domain for $\tau\in{\cal H}_1$ is
determined, the full genus two fundamental domain ${\cal F}_2'$ will
follow from the other transformation rules in (\ref{sl2subgroup}).

\subsection{Moduli Space}

We will now construct a fundamental modular domain in the upper
complex half-plane ${\cal H}_1$ for the action of the congruence subgroup
$\Gamma_{(\mu)}\subset SL(2,\zed)$. Let
\beq
S=\left(\begin{array}{cc}0&-1 \\ 1&0\end{array}\right) \ , \quad
T=\left(\begin{array}{cc}1&1 \\ 0&1\end{array}\right)
\label{SL2Zgens}\eeq
be the standard generators of $SL(2,\zed)$. Consider the fundamental
domain $\Delta$ for the action of $SL(2,\zed)$ given by
(\ref{calF1}), which is a triangle with one vertex at infinity. The
three edges separate $\Delta$ from the M\"obius images $S\sbullet\Delta$,
$T\sbullet\Delta$ and $T^{-1}\sbullet\Delta$. A {\it Schreier transversal}
$\mathfrak{C}_{(\mu)}$ for $\Gamma_{(\mu)}$ in $SL(2,\zed)$ with
respect to $\{S,T\}$ is a set of right coset representatives
$SL(2,\zed)=\bigcup_g\,\Gamma_{(\mu)}g$
(i.e. $\Gamma_{(\mu)}g\cap\mathfrak{C}_{(\mu)}$ has precisely one
element for each $g\in SL(2,\zed)$) expressed as
words in the generating set $\{S,T\}$ such that each prefix (or initial
segment) of an element of $\mathfrak{C}_{(\mu)}$ is also in
$\mathfrak{C}_{(\mu)}$. Then the region
\begin{equation}\label{schreier}\mathfrak{C}_{(\mu)}\sbullet\Delta
=\bigcup_{C\in\mathfrak{C}_{(\mu)}}\,C\sbullet\Delta\end{equation}
is a polygonal fundamental domain for the action of $\Gamma_{(\mu)}$
on ${\cal H}_1$. For example, if $S\,T\,S\in\mathfrak{C}_{(\mu)}$,
then also $S\,T,S,\id_2\in\mathfrak{C}_{(\mu)}$. The triangles
$(S\,T\,S)\sbullet\Delta$ and $(S\,T)\sbullet\Delta$ share a common
edge, as do $(S\,T)\sbullet\Delta$ and $S\sbullet\Delta$, and so on.

Since the subgroup $\Gamma_{(\mu)}\subset SL(2,\zed)$ has finite
index, there are finite Schreier transversals. The group
$\Gamma_{(\mu)}$ is the preimage of the subgroup
\begin{equation}\label{gammasub}
\phi\big(\Gamma_{(\mu)}\big)=
\tilde\Gamma_{(\mu)}:=\Bigg\{\left(\begin{array}{cc}\mu\,a+1&b \\ 0 &
    \mu\,d+1\end{array}\right)\quad\Bigg|\quad
a+d\equiv0~{\rm mod}\,\mu \ , \ b\in\zed/\mu^2\,\zed\Bigg\}
\end{equation}
of the finite group $SL(2,\zed/\mu^2\,\zed)$ under the
surjective homomorphism
\beq
\phi\,:\,SL(2,\zed)~\longrightarrow~SL(2,\zed/\mu^2\,\zed)
\eeq
given by reduction modulo $\mu^2$. The index of $\Gamma_{(\mu)}$ in
$SL(2,\zed)$ may thereby be computed from
\bea
\bigl[SL(2,\zed):\Gamma_{(\mu)}\bigr]&=&\bigl[{\rm im}(\phi):
\tilde\Gamma_{(\mu)}\cap{\rm im}(\phi)\bigr]\nonumber\\
&=&\bigl[SL(2,\zed/\mu^2\,\zed):\tilde\Gamma_{(\mu)}\bigr]~=~
\frac{\bigl|SL(2,\zed/\mu^2\,\zed)\bigr|}{\bigl|\tilde\Gamma_{(\mu)}
\bigr|} \ . \eea
We now need to work out the orders of the two finite groups
$SL(2,\zed/\mu^2\,\zed)$ and $\tilde\Gamma_{(\mu)}$. The order of
$\tilde\Gamma_{(\mu)}$ can be easily determined by inspection of its
definition (\ref{gammasub}) to be
$|\tilde\Gamma_{(\mu)}|=\mu^3$. The order of
$SL(2,\zed/\mu^2\,\zed)$ is calculated as follows.

The index of $\Gamma_{(\mu)}$ turns out to depend crucially on the
prime factorization of the integer $\mu$. Suppose that
$\mu=p^{k(1)}_1\cdots p_t^{k(t)}$ with $p_j$, $j=1,\dots,t$ distinct
prime numbers and $k(j)>0$. By the Chinese remainder theorem the
corresponding finite group factorizes as
\beq
SL\bigl(2,\zed/\mu^2\,\zed\bigr)=SL\bigl(2,\zed/p^{2k(1)}_1\,\zed
\bigr)~\times~\cdots~\times~
SL\bigl(2,\zed/p^{2k(t)}_t\,\zed\bigr)
\label{Chinese}\eeq
and its order is given by
\beq
\bigl|SL(2,\zed/\mu^2\,\zed)\bigr|=\prod^t_{j=1}\,\bigl
|SL(2,\zed/p^{2k(j)}_j\,\zed)\bigr| \ .
\eeq
It thus suffices to compute the order of $SL(2,\zed/p^{2k}\,\zed)$ for
$p$ prime and $k>0$. Let $\big({}^a_c\,{}^b_d\big)\in
SL(2,\zed/p^{2k}\,\zed)$. Then $a\,d-b\,c\equiv
1~{\rm{mod}}\,p^{2k}$. To ensure that the matrix is non-singular,
the pair $(a,b)$ must take values in the set
\beq
\bigl(\zed/p^{2k}\,\zed~\times~\zed/p^{2k}\,\zed\bigr)\,
\setminus\,\bigl(p\,\zed/p^{2k}\,\zed~\times~
p\,\zed/p^{2k}\,\zed\bigr) \ .
\label{absetvalue}\eeq
The number of elements in this set is $p^{4k-2}\,(p^2-1)$. The
pair $(c,d)$ must be chosen so that $p$ does not divide the
determinant. There are $p^{4k-1}\,(p-1)$ such pairs $(c,d)$ for
each $(a,b)$. This ensures that the matrix is non-singular. Thus the
number of invertible matrices is given by
\beq
\bigl|GL(2,\zed/p^{2k}\,\zed)\bigr|=p^{4k-2}\,\bigl(p^2-1
\bigr)\,p^{4k-1}\,\bigl(p-1\bigr) \ .
\eeq
The determinant is a group homomorphism
$\det:GL(2,\zed/p^{2k}\,\zed)\to\zed/p^{2k}\,\zed$. It
follows that the index of $SL(2,\zed/p^{2k}\,\zed)$ in
$GL(2,\zed/p^{2k}\zed)$ is
\beq
\bigl[GL(2,\zed/p^{2k}\,\zed):SL(2,\zed/p^{2k}\,\zed)
\bigr]=\frac{\bigl|GL(2,\zed/p^{2k}\,\zed)\bigr|}
{\bigl|SL(2,\zed/p^{2k}\,\zed)\bigr|}=p^{2k-1}\,(p-1) \ ,
\eeq
which is just the Euler $\varphi$-function of the field
$\zed/p^{2k}\,\zed$.

\begin{figure}[hbt]
\begin{center}
\epsfxsize=5.0 in\epsfbox{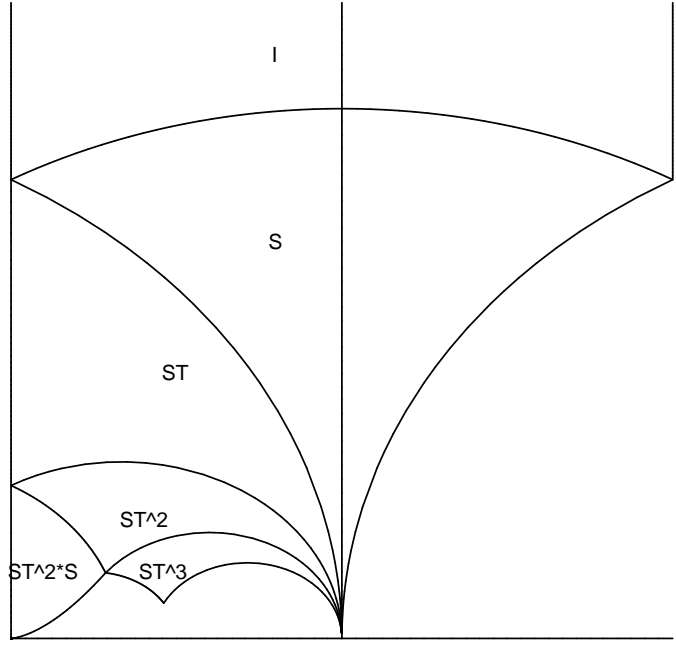}
\end{center}
\caption{The fundamental domain $\mathfrak{C}_{(2)}\sbullet\Delta$ for the
  action of the congruence subgroup $\Gamma_{(2)}\subset SL(2,\zed)$
  on ${\cal H}_1$.}
\label{Schreier2}\end{figure}

By combining all of these results we find finally that the index of
$\Gamma_{(\mu)}$ in $SL(2,\zed)$ is given by the Euler product
expansion
\be
\bigl[SL(2,\zed):\Gamma_{(\mu)}\bigr]=
\mu^3\,\prod_{\textrm{primes}\,p|\mu}\,\left(1-\frac1{p^2}\right) \ .
\label{Gammaindex}\ee
We can now build a Schreier tranversal inductively, starting from
$\{\id_2\}$. Suppose that we have a set $\mathfrak{C}_k$ of $k$ words,
satisfying the suffix condition, which contains at most one
representative of any right coset.  If $k$ is strictly less than the
index (\ref{Gammaindex}), then we can examine the right cosets
$\Gamma_{(\mu)}S\,C$, $\Gamma_{(\mu)}T\,C$ and
$\Gamma_{(\mu)}T^{-1}\,C$ for each $C\in\mathfrak{C}_k$ until we find
one which is different from $\Gamma_{(\mu)}C$ for
$C\in\mathfrak{C}_k$. Then add $S\,C$, $T\,C$ or $T^{-1}\,C$ to the
list of words to form a new list $\mathfrak{C}_{k+1}$. This process
terminates precisely when $k$ is equal to the index
(\ref{Gammaindex}), and then $\mathfrak{C}_k=\mathfrak{C}_{(\mu)}$ is the
desired Schreier transversal for $\Gamma_{(\mu)}$. For example, when
$\mu=2$ the subgroup $\Gamma_{(2)}$ has index $6$ and
$\mathfrak{C}_{(2)}=\{\id_2,S,S\,T,S\,T^2,S\,T^3,S\,T^2\,S\}$ is a Schreier
transversal for $\Gamma_{(2)}$. The corresponding elliptic fundamental
domain (\ref{schreier}) is depicted in Figure~\ref{Schreier2}.

Using the modular transformations (\ref{sl2subgroup}) along with the
positivity constraint (\ref{Omega2pos}) on the period matrix, we find
that the fundamental domain at genus two for the residual modular group
preserving the Poincar\'e normal form is given by
\begin{equation}
{\cal F}'_{2}(\mu)=\left(\mathfrak{C}_{(\mu)}\sbullet\Delta\right)
\times\complex\times{\cal H}_1
\end{equation}
with elements $(\Omega_{22},\Omega_{12},\Omega_{11})$. The integers
$\mu,t,r,a,b,c$ and $d$ are thus unrestricted except for the
dependences of $a,b,c$ and $d$ on $r,s$ and $t$. Because of
this dependence and the complexity of the integration region
$\tau\in\mathfrak{C}_{(\mu)}\sbullet\Delta$, the free energy cannot be
made as explicit as those computed in Sections~\ref{BosFreeen},
\ref{SUSYFreeen} and \ref{Heterotic}.

\setcounter{subsection}{0}
\setcounter{subsubsection}{0}

\appendix{Explicit Form of $\mbf{\Xi_6}$}

In this appendix we provide the explicit expressions for the modular
covariant form $\Xi_6[\delta]$ on ${\cal H}_2$ defined in
(\ref{Xi6def}) for the ten even spin structures. Given an even
characteristic $\delta_i$, $i=0,1,\dots,9$, we denote
$\vartheta_i(\Omega):=\Theta[\delta_i](\mbf 0|\Omega)^4$. By the
mirror property~\cite{DHPh1}, there are two equivalent expressions for
$\Xi_6[\delta_i](\Omega)$ corresponding to the two triples of odd spin
structures used to represent
$\delta_i=\nu_{i_1}+\nu_{i_2}+\nu_{i_3}=\nu_{i_4}+\nu_{i_5}+\nu_{i_6}$
for each $i$. One then has
\bea
\Xi_6[\delta_1]&=&-\vartheta_4\,\vartheta_5\,\vartheta_8-\vartheta_2\,
\vartheta_6\,\vartheta_9-\vartheta_3\,\vartheta_7\,\vartheta_0~=~
-\vartheta_4\,\vartheta_7\,\vartheta_6-\vartheta_3\,\vartheta_8\,
\vartheta_9-\vartheta_2\,\vartheta_5\,\vartheta_0 \ , \nonumber\\[5pt]
\Xi_6[\delta_2]&=&\vartheta_3\,\vartheta_5\,\vartheta_7+\vartheta_4\,
\vartheta_8\,\vartheta_0-\vartheta_1\,\vartheta_6\,\vartheta_9~=~
\vartheta_3\,\vartheta_6\,\vartheta_8-\vartheta_1\,\vartheta_5\,
\vartheta_0+\vartheta_4\,\vartheta_7\,\vartheta_9 \ , \nonumber\\[5pt]
\Xi_6[\delta_3]&=&\vartheta_2\,\vartheta_5\,\vartheta_7-\vartheta_1\,
\vartheta_8\,\vartheta_9+\vartheta_4\,\vartheta_6\,\vartheta_0~=~
\vartheta_2\,\vartheta_6\,\vartheta_8+\vartheta_5\,\vartheta_4\,
\vartheta_9-\vartheta_1\,\vartheta_7\,\vartheta_0 \ , \nonumber\\[5pt]
\Xi_6[\delta_4]&=&-\vartheta_1\,\vartheta_5\,\vartheta_8+\vartheta_3\,
\vartheta_6\,\vartheta_0+\vartheta_2\,\vartheta_7\,\vartheta_9~=~
-\vartheta_1\,\vartheta_6\,\vartheta_7+\vartheta_2\,\vartheta_8\,
\vartheta_0+\vartheta_3\,\vartheta_5\,\vartheta_9 \ , \nonumber\\[5pt]
\Xi_6[\delta_5]&=&\vartheta_2\,\vartheta_3\,\vartheta_7-\vartheta_1\,
\vartheta_4\,\vartheta_6+\vartheta_6\,\vartheta_9\,\vartheta_0~=~
-\vartheta_1\,\vartheta_2\,\vartheta_0+\vartheta_3\,\vartheta_4\,
\vartheta_9+\vartheta_6\,\vartheta_7\,\vartheta_8 \ , \nonumber\\[5pt]
\Xi_6[\delta_6]&=&\vartheta_3\,\vartheta_4\,\vartheta_0-\,\vartheta_1\,
\vartheta_2\,\vartheta_9+\vartheta_5\,\vartheta_7\,\vartheta_8~=~
-\vartheta_1\,\vartheta_4\,\vartheta_7+\vartheta_5\,\vartheta_9\,
\vartheta_0+\vartheta_2\,\vartheta_3\,\vartheta_8 \ , \nonumber\\[5pt]
\Xi_6[\delta_7]&=&\vartheta_2\,\vartheta_3\,\vartheta_5+\vartheta_8\,
\vartheta_9\,\vartheta_0-\vartheta_1\,\vartheta_4\,\vartheta_6~=~
\vartheta_2\,\vartheta_4\,\vartheta_9-\vartheta_1\,\vartheta_3\,
\vartheta_0+\vartheta_5\,\vartheta_6\,\vartheta_8 \ , \nonumber\\[5pt]
\Xi_6[\delta_8]&=&\vartheta_7\,\vartheta_9\,\vartheta_0-\vartheta_1\,
\vartheta_4\,\vartheta_5+\vartheta_2\,\vartheta_3\,\vartheta_6~=~
-\vartheta_1\,\vartheta_3\,\vartheta_9+\vartheta_2\,\vartheta_4\,
\vartheta_0+\vartheta_5\,\vartheta_6\,\vartheta_7 \ , \nonumber\\[5pt]
\Xi_6[\delta_9]&=&\vartheta_7\,\vartheta_8\,\vartheta_0-\vartheta_1\,
\vartheta_2\,\vartheta_6+\vartheta_3\,\vartheta_4\,\vartheta_5~=~
\vartheta_5\,\vartheta_6\,\vartheta_0-\vartheta_1\,\vartheta_3\,
\vartheta_8+\vartheta_2\,\vartheta_4\,\vartheta_7 \ , \nonumber\\[5pt]
\Xi_6[\delta_0]&=&\vartheta_7\,\vartheta_8\,\vartheta_9+\vartheta_3\,
\vartheta_4\,\vartheta_6-\vartheta_1\,\vartheta_2\,\vartheta_5~=~
\vartheta_5\,\vartheta_6\,\vartheta_9+\vartheta_2\,\vartheta_4\,
\vartheta_8-\vartheta_1\,\vartheta_3\,\vartheta_7 \ .
\eea

\end{document}